# Superhabitable Worlds

René Heller[1], John Armstrong[2]

[1] McMaster University, Department of Physics and Astronomy, Hamilton, ON L8S 4M1, Canada (rheller@physics.mcmaster.ca)
[2] Department of Physics, Weber State University, 2508 University Circle, Ogden, UT 84408-2508 (jcarmstrong@weber.edu)

To be habitable, a world (planet or moon) does not need to be located in the stellar habitable zone (HZ), and worlds in the HZ are not necessarily habitable. Here, we illustrate how tidal heating can render terrestrial or icy worlds habitable beyond the stellar HZ. Scientists have developed a language that neglects the possible existence of worlds that offer more benign environments to life than Earth does. We call these objects "superhabitable" and discuss in which contexts this term could be used, that is to say, which worlds tend to be more habitable than Earth. In an appendix, we show why the principle of mediocracy cannot be used to logically explain why Earth should be a particularly habitable planet or why other inhabited worlds should be Earth-like.

Superhabitable worlds must be considered for future follow-up observations of signs of extraterrestrial life. Considering a range of physical effects, we conclude that they will tend to be slightly older and more massive than Earth and that their host stars will likely be K dwarfs. This makes Alpha Centauri B, member of the closest stellar system to the Sun that is supposed to host an Earth-mass planet, an ideal target for searches of a superhabitable world.

**Key Words:** Extrasolar terrestrial Planets – Extraterrestrial Life – Habitability – Planetary environments – Tides

## 1. Introduction

A substantial amount of research is conducted and resources are spent to search for planets that could be habitats for extrasolar life. Engineers and astronomers have developed expensive instruments and large ground-based telescopes, such as the *High Accuracy Radial Velocity Spectrograph* (HARPS) at the 3.6m ESO telescope and the *Ultraviolet-Visual Echelle Spectrograph* (UVES) at the *Very Large Telescope* (VLT), and launched the *CoRoT* and *Kepler* space telescopes with the explicit aim to detect and characterize Earth-sized planets. Even larger facilities are being planned or constructed, such as the *European Extremely Large Telescope* (E-ELT) and the *James Webb Space Telescope* (JWST), and an ever growing community of scientists is working to solve not only the observational but also the theoretical and laboratory challenges.

At the theoretical front, the concept of the stellar "habitable zone" (HZ) has been widely used to identify potentially habitable planets (Huang, 1959; Dole, 1964; Kasting et al., 1993). To the confusion of some, planets that reside within a star's HZ are often called "habitable planets." However, a planet in the HZ need not be habitable in the sense that it has at least some niches that allow for the existence of liquid surface water. Naturally, as Earth is the only inhabited world we know, this object usually serves as a reference for studies on habitability. Instruments are being designed in a way to detect and characterize Earth-like planets and spectroscopic signatures of life in Earth-like atmospheres (Des Marais et al., 2002; Kaltenegger and Traub, 2009; Kaltenegger et al., 2010; Rauer et al., 2011; Kawahara et al., 2012). However, other worlds can offer conditions that are even more suitable for life to emerge and to evolve. Besides planets, moons could be habitable, too (Reynolds et al., 1987; Williams et al., 1997; Kaltenegger, 2010; Porter and Grundy, 2011; Heller and Barnes, 2013a). To find a habitable and ultimately an inhabited world, a characterization concept is required that is biocentric rather than geo- or anthropocentric.

In Section 2 of this paper, we illustrate how tidal heating can make planets inside the stellar HZ uninhabitable, and how it can render exoplanets and exomoons beyond the HZ habitable. Section 3 is devoted to conditions that could make a world more hospitable for life than Earth is. We call these objects "superhabitable worlds." Though our considerations are anticipatory, they still rely on the assumption that life needs liquid water. Our conclusions on the nature and prospects for finding superhabitable worlds are presented in Section 4. In Appendix A, we disentangle confusions between planets in the HZ and habitable planets, and we address related disorder that emerges from language issues. Appendix B is dedicated to the principle of mediocracy – in particular why it cannot explain that Earth is a typical, inhabited world.

## 2. Habitability in and Beyond the Stellar Habitable Zone

### *2.1 The stellar habitable zone*

A natural starting point towards the characterization of a world's habitability is computing its absorbed stellar energy flux. This approach has led to what is called the "stellar habitable zone." The oldest record of a description of a circumstellar zone suitable for life traces back to Whewell (1853, Chap. X, Section 4), who, referring to the local stellar system in a qualitative way, called this distance range the "Temperate Zone of the Solar System." More than a century later, Huang (1959) presented a more general discussion of the "Habitable Zone of a Star," which considers time scales of stellar evolution, dynamical constraints in stellar multiple systems, and the stellar galactic orbit. A much broader, less anthropocentric elaboration on habitability has then been given by Dole (1964), who termed the circumstellar habitable zone "ecosphere."[1] The most widely used concept as of today is the one presented by Kasting et al. (1993), who applied a one-dimensional climate model and identified the $CO_2$ feedback to ensure the inner and the outer edges of the stellar HZs. The inner edge is defined by the activation of the moist or runaway greenhouse process, which desiccates the planet by evaporation of atmospheric hydrogen; the outer edge is defined by $CO_2$ freeze out, which breaks down the greenhouse effect whereupon the planet transitions into a permanent snowball state. Extensions of this concept to include orbital eccentricities have been given by Selsis et al. (2007) and Barnes et al. (2008), and a recent revision of the input model used by Kasting et al. (1993) has been presented by Kopparapu et al. (2013).

Considering the aging of the star, which involves a steady increase of stellar luminosity as long as the star is on the main sequence, the distance range within the HZ that is habitable for a certain period (say over the last 4.6 Gyr in the case of the Solar System) has been termed the continuous habitable zone (CHZ) (Kasting et al., 1993; Rushby et al., 2013). From an observational point of view, the CHZ provides a more useful tool because life needs time to evolve to a certain level such that it modifies its atmosphere on a global scale. Life seems to have appeared relatively early after the formation of Earth. Chemical and fossil indicators for early life can be found in sediments that date back to about 3.5 – 3.8 Gyr ago (Schopf, 1993; Mojzsis et al., 1996; Schopf, 2006; Basier et al., 2006), that is, less than about 1 Gyr after Earth had formed. If correct, then life would have recovered within 100 Myr or so after the Late Heavy Bombardment (LHB) on Earth (Gomes et al., 2005). However, life required billions of years before it modified Earth's atmosphere substantially and imprinted substantial amounts of bio-relevant signatures in the atmospheric transmission spectrum. Stars more massive than the Sun have shorter lifetimes. Thus, although the lifetime of a 1.4 solar-mass ($M_\odot$) star is still about 4.5 Gyr, superhabitable planets will tend to orbit stars that are as massive as the Sun at most.

Further modifications of the circumstellar HZ include effects of tidal heating (Jackson et al., 2008a; Barnes et al., 2009), orbital evolution due to tides (Barnes et al., 2008), tidal locking of planetary rotation (Dole, 1964; Kasting et al., 1993), planetary obliquity (Spiegel et al., 2009), loss of seasons due to tilt erosion (Heller et al., 2011), land-to-ocean fractional coverage on planets (Spiegel et al., 2008), stellar irradiation in eccentric orbits (Dressing et al., 2010; Spiegel et al., 2010), the formation of water clouds on tidally locked planets (Yang et al., 2013), and the dependence of the ice-albedo feedback on the stellar spectrum and the planetary atmosphere (Joshi and Haberle, 2012; von Paris et al., 2013). These studies show that planets in the HZ of stars with masses $M_* \lesssim 0.5\ M_\odot$ can be subject to enormous tidal heating, substantial variations in their semi-major axis, loss of seasons, and tidal locking. Above all, they demonstrate that the circumstellar HZ, though a helpful working concept, does not define a planet's habitability. For one and the same star, two different planets can have a different HZ, depending on a myriad of bodily and orbital characteristics.

### *2.2 A terrestrial menagerie*

Accounting for some of these effects, we can imagine a menagerie of terrestrial worlds. In Fig. 1, we show these planets, all of which are assumed to have a mass 1.5 times that of Earth ($M_p = 1.5\ M_\oplus$), a radius 1.12 that of Earth[2], and a host star similar to Gl581 (Mayor et al., 2009). Rather than discussing the exact orbital limits for any of these hypothetical worlds, we shall illustrate here the range of possible scenarios. Irradiation from the star is given by $F_i$, tidal heat flux by $F_t$ (computed with the Leconte et al., 2010 tidal equilibrium model), and the critical flux for the planet to initiate a runaway greenhouse effect by $F_{RG}$ (Goldblatt and Watson, 2012). Using the analytical expression given in Pierrehumbert (2010), we estimate $F_{RG}$ = 301 W/m$^2$ for our test planet, and we compute the HZ boundaries with the model of Kopparapu et al. (2013). Following the approach of Barnes et al. (2013) and Heller and Barnes (2013b), we identify the following members of the menagerie:

- **Tidal Venus:** $F_t \geq F_{RG}$ (Barnes et al., 2013)
- **Insolation Venus:** $F_i \geq F_{RG}$
- **Tidal-Insolation Venus:** $F_t < F_{RG}$, $F_i < F_{RG}$, $F_t + F_i \geq F_{RG}$
- **Super-Io:** $F_t > 2$ W/m$^2$, $F_t + F_i < F_{RG}$ (hypothesized by Jackson et al., 2008b)
- **Tidal Earth:** 0.04 W/m$^2$ $< F_t <$ 2 W/m$^2$, $F_t + F_i < F_{RG}$ and within the HZ
- **Super-Europa:** 0.04 W/m$^2$ $< F_t <$ 2 W/m$^2$ and beyond the HZ
- **Earth twin:** $F_t <$ 0.04 W/m$^2$ and within the HZ
- **Snowball Earth:** $F_t <$ 0.04 W/m$^2$ and beyond the HZ

Among these worlds, a Tidal Venus, an Insolation Venus, and a Tidal-Insolation Venus are uninhabitable by definition, while a Super-Io, Tidal Earth, Super-Europa, and an Earth Twin could be habitable. The surface of a Snowball Earth is also uninhabitable because it is so cold that even atmospheric $CO_2$ would condense, and the warming greenhouse effect could not operate to maintain liquid surface water. Note that the 2 and 0.04 W/m$^2$ thresholds are taken from the Solar System, where it has been observed that Io's global volcanism coincides with a surface flux of 2 W/m$^2$ (Spencer et al., 2000). Moreover, Williams et al. (1997) suggested that tectonic activity on Mars ceased when its endogenic surface flux fell below 0.04 W/m$^2$. Concerning the Super-Europa class, note that O'Brien et al. (2002) estimated Europa's tidal heat flux to about 0.8 W/m$^2$.

[1] Dole notes that the term "ecosphere" goes back to Strughold (1955).

[2] The radius is derived with an assumed Earth-like rock-to-mass fraction of 0.68 and using the analytical expression provided by Fortney et al. (2007).

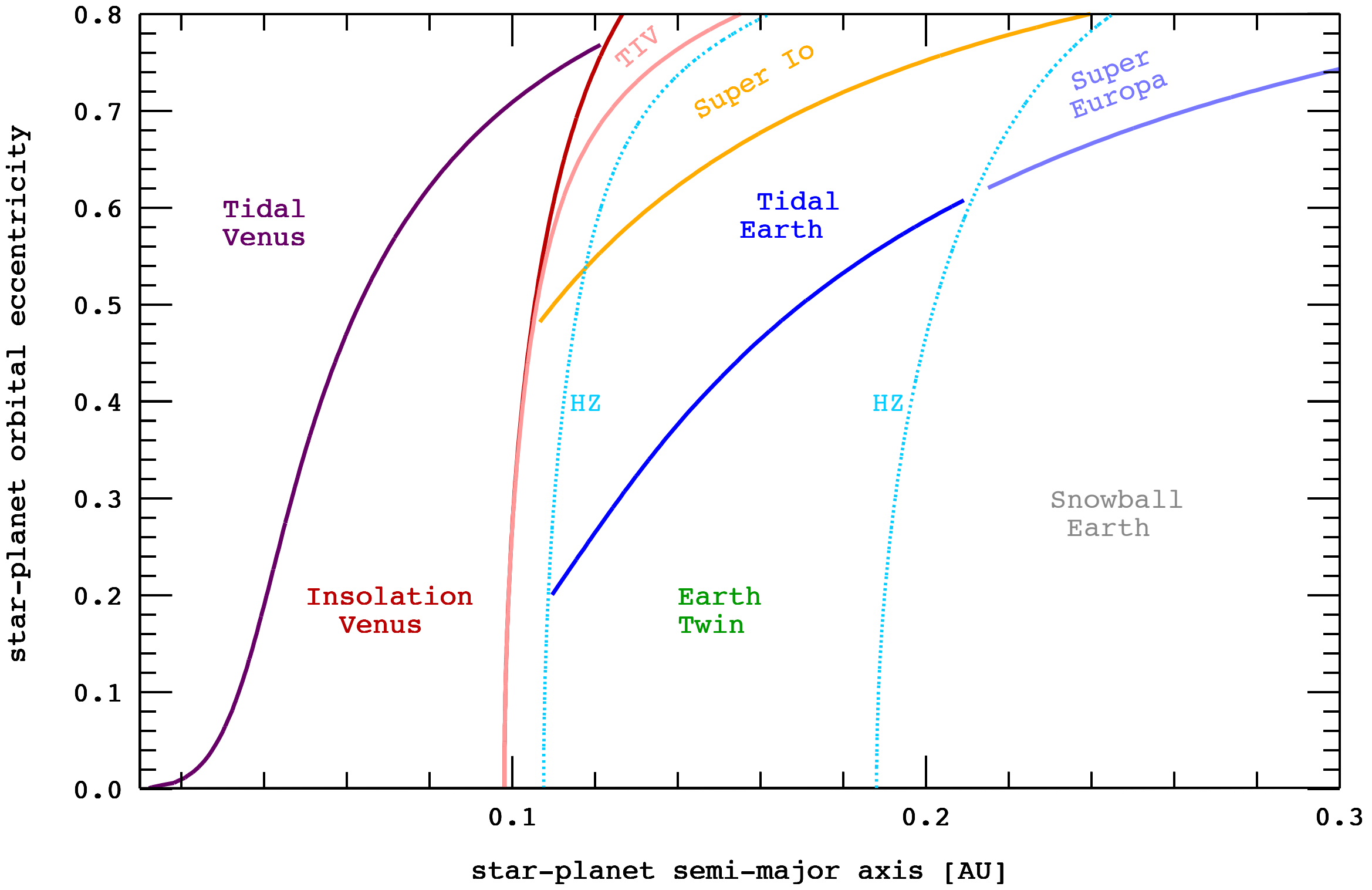

**Fig. 1**: Menagerie of terrestrial planets based on stellar irradiation and tidal heating. A planet with a mass of 1.5 $M_\oplus$ in orbit around a star similar to Gl581 is assumed. Dashed lines indicate the boarders of the HZ following Kopparapu et al. (2013). The inner edge is constituted by the runaway greenhouse effect, the outer limit by the maximum greenhouse effect. Note that tidal heating can potentially heat planets beyond the HZ and open a class of Super Europas.

This menagerie illustrates that terrestrial planets can be located in the HZ and yet be uninhabitable. Tidal heating during the planet's orbital circularization can be an additional heat source that causes a planet to enter a runaway greenhouse state. What is more, tidal heating could make a world habitable *beyond* the HZ, possibly the Super-Europa planets in our menagerie. Elevated orbital eccentricities would induce tidal friction in these planets, which would transform orbital energy into heat. Such highly eccentric orbits would tend to be circularized, and hence perturbations from other planets or stars in the system would be required to maintain substantial eccentricities. Then tidal heating could partly compensate for the reduced stellar illumination beyond the stellar HZ and potentially maintain liquid water reservoirs.

### *2.3 Habitable exomoons beyond the stellar habitable zone*

In exomoons beyond the stellar HZ, tidal heat could even become the major source of energy to allow for liquid water – be it on the surface or below (Reynolds et al., 1987; Scharf, 2006; Debes and Sigurdsson, 2007; Cassidy et al., 2009; Henning et al., 2009; Heller and Barnes, 2013a,b). Imagine a moon the size and mass of Earth in orbit around a planet the size and mass of Jupiter, and assume that this binary orbits a star of solar luminosity at a distance of 1 AU. If the moon is in a wide orbit, say beyond 20 planetary radii from its host, then it will hardly receive stellar reflected light or thermal emission from the planet (Heller and Barnes, 2013a,b), its orbit-averaged stellar illumination will not be substantially reduced by eclipses behind the planet (Heller, 2012), tidal heating will be insignificant[3], and the moon will essentially be heated by illumination absorbed from the star. But as the moon is virtually shifted into a closer orbit around the planet, illumination from the planet and tidal heating increase, and the total energy flux can become large enough to render the moon uninhabitable. The critical orbit, in which the total energy flux equals the critical flux for the moon to enter the runaway greenhouse effect, has been termed the circumplanetary "habitable edge" (Heller and Barnes, 2013a). Moons inside the habitable edge are uninhabitable. Imagine further that the planet-moon binary is virtually shifted away from the star. Due to the reduced stellar illumination, the habitable edge moves inward towards the planet because tidal heat and illumination from the planet can outbalance the loss of stellar illumination. When the planet-moon system is shifted even beyond the stellar HZ, then the moon will *need* to be close enough to the planet such that it will prevent transition into a snowball state. In this sense, giant planets beyond the stellar HZ have their own circumplanetary HZ, defined by illumination from the planet and tidal heating in the moon.

[3] Only if the moon's rotation is fast after formation, then it can experience tidal heating in a wide orbit due to the deceleration towards synchronous rotation. In this particular constellation of an Earth-like moon around a Jupiter-like planet at 1 AU from a Sun-like star, this tidal locking takes less than 4.5 Gyr, even in the widest possible orbits (Hinkel and Kane, 2013). Moreover, tidal heating can be substantial even beyond 20 planetary radii from the host planet if the moon's orbital eccentricity is large. However, circularization will damp it within a few Myr (Porter and Grundy, 2011; Heller and Barnes, 2013a).

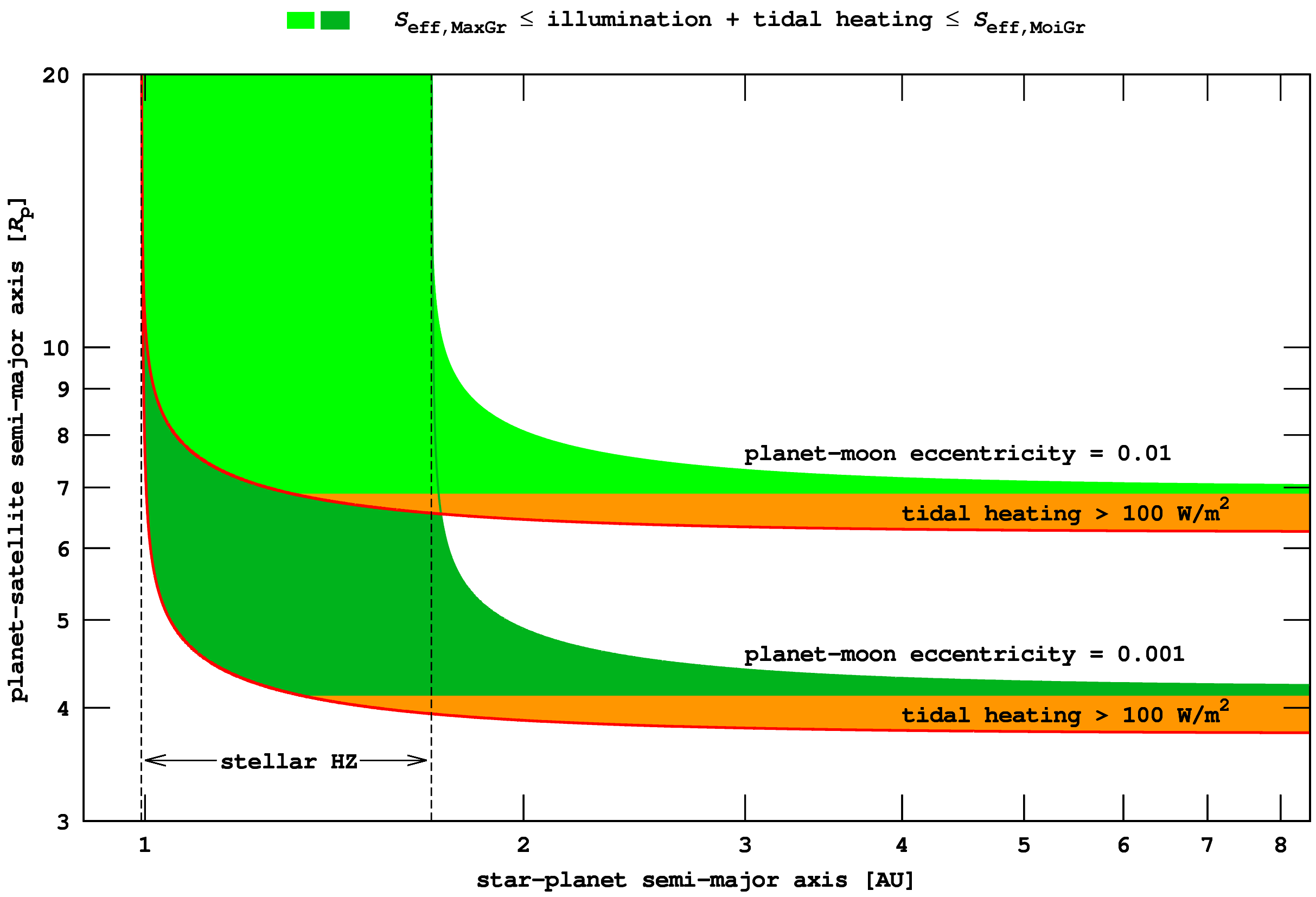

**Fig. 2:** Habitable orbits for an Earth-like exomoon around a Jupiter-like planet around a solar luminosity star. Green areas ilustrate orbits, in which the total energy flux of absorbed illumination and tidal heating is above the maximum greenhouse limit $S_{\mathrm{eff,MaxGr}}$ and below the moist greenhouse threshold $S_{\mathrm{eff,MoiGr}}$. Within these stripes, orbits with tidal heating rates above 100 W/m$^2$ are highlighted in orange. The circumplanetary habitable edge, here defined by the moist greenhouse, is indicated with a red line.

In Fig. 2, we illustrate this scenario for two different orbital eccentricities of the planet-moon binary, 0.01 and 0.001. The abscissa denotes stellar distance of the planet-moon system, and the ordinate shows the distance between the planet and its satellite. Green areas denote orbits, in which the total flux – composed of stellar plus planetary illumination and tidal heating – varies between the minimum and maximum energy flux ($S_{\mathrm{eff,MaxGr}}$ and $S_{\mathrm{eff,MoiGr}}$, respectively) identified by Kopparapu et al. (2013) to define the solar HZ. To compute the total energy flux, we chose the same model as in Heller and Barnes (2013a)[4]. As the planet-moon system is assumed at increasing stellar distances, the habitable edge (red line) moves closer to the planet. Ultimately, beyond the stellar HZ, the satellite must be closer to its planet than a certain maximum distance such that it receives enough tidal heating. Moving to the outer regions of the star system, stellar irradiation vanishes and tidal heat becomes the dominant source of energy. Comparison of the two stripes in Fig. 2 indicates that moons in orbits with only small orbital eccentricities would need to be closer to the planet to experience substantial tidal heating.

Note that the orbital eccentricities of the Galilean satellites around Jupiter are all larger than 0.001, and that Titan's eccentricity around Saturn is 0.0288. While the reason for Titan's enhanced eccentricity remains unclear (Sohl et al., 1995), the eccentricities of the major Jovian moons are not free but forced, that is, they are excited by the satellites' gravitational interaction (Yoder, 1979). Thus, as rocky and icy exomoons are predicted to exist around extrasolar Jovian planets (Sasaki et al., 2010; Ogihara and Ida, 2012), and if these moons encounter substantial orbital perturbations by other moons, then possibly many habitable exomoons in and beyond the stellar HZ await their discovery.

[4] This model, which includes a tidal theory presented by Leconte et al. (2010), neglects the feedback between tidal heating and the rheology of the moon. Yet, it has been shown that increasing tidal heat can melt a terrestrial body, thereby shutting down tidal heating itself (Zahnle et al., 2007; Henning et al., 2009; Remus et al., 2012).

## 3. Physical Characteristics of Superhabitable Worlds

All exoplanets detected so far are either subject to stellar irradiation that is very different from the amount or spectral distribution currently received by Earth, or they have masses larger than a few Earth masses. This has led astrobiologists to speculate about extremophile life forms that could cope with more bizarre conditions and maybe survive on a planet that is more hostile than Earth (for a brief review see Dartnell, 2011). The word "bizarre" is here to be understood from an anthropocentric point of view. From a potpourri of habitable worlds that may exist, Earth might well turn out as one that is marginally habitable[5], eventually bizarre from a biocentric standpoint. In other words, it is not clear why Earth should offer the most suitable regions in the physicochemical parameter space that can be tolerated by living organisms. Such an anthropocentric assumption could mislead research for extrasolar habitable planets because planets could be non-Earth-like but yet offer more suitable conditions for the emergence and evolution of life than Earth did or does, that is, they could be superhabitable.

As to how superhabitable planets could look like or under which conditions a world could occupy a more benign zone within the physicochemical volume, we now discuss planetary characteristics that are relevant to planetary habitability (for a broader review see Gaidos et al., 2005; Lammer et al., 2009). These considerations will allow us to deduce quantitative estimates for superhabitable worlds. Instead of elaborating on extremophile or even completely different forms of life, we will still stick to liquid water as a pre-requisite for life and explore more comfortable environments as those found on Earth. Thus, our extensions of habitability towards superhabitability are incremental and still carry a geocentric flavor.

What could we understand under a superhabitable world? So far, the term has not been in use, and thus its meaning remains obscure. We propose a context family in which it might be used with reason.[6]

- Habitable surface area: An Earth-sized planet on which the surface area that permits liquid water is larger than that of Earth (Spiegel et al., 2009; Pierrehumbert, 2010, § 1.9.1) could be regarded as superhabitable.

- Total surface area: A more uneven surface, or simply a larger planet with more space for living forms, could make a planet superhabitable. Due to the higher surface gravity of a more massive planet, however, both characteristics tend to exclude one another. To increase a planet's habitability, the body cannot be arbitrarily large. As mass typically increases with increasing radius for terrestrial planets, plate tectonics will cease to operate at a certain mass (see below). Moreover, a terrestrial planet that is much heavier than Earth might not get rid of its primordial hydrogen atmosphere, which could hamper the emergence of life (Huang, 1960). A planet slightly larger than Earth can, however, still be regarded superhabitable. Note that Earth, the only inhabited planet known so far, is the largest terrestrial planet in the Solar System.

- Land-to-ocean-fraction and distribution: The amount of surface water compared to the amount of land is not only crucial for planetary climate but also for the emergence and diversification of life. Giant continents, as Earth's Gondwana about 500 Myr ago, may have vast deserts in their interiors, as they are not subject to the moderating effect of oceans. In contrast, planets with more fractionate continents and archipelagos should favor superhabitable environments due to their enhanced richness in habitats. Earth's shallow waters have a higher biodiversity than the deep oceans (Gray, 1997). Hence, we expect that planets with shallow waters rather than those with deep extended oceans tend to be superhabitable.

  What is more, Abe et al. (2011) found that planets dryer than Earth should have wider stellar HZs. At the inner HZ boundary, these "Dune" planets are more tolerant against transition into the runaway greenhouse effect, because their low-humidity equatorial regions can emit above the critical flux, assumed for an atmosphere saturated in water. But still the atmosphere is somewhat opaque in the infrared and thus exerts a global greenhouse effect, which prevents water at poles from freezing. On dry planets at the outer HZ boundary, the low humidity in the tropics hampers formation of clouds and thus snowfall. Dry planets will thus tend to have lower albedos than frozen aqua planets (such as Earth), and they will effectively absorb more stellar illumination and be less susceptible to transitioning into a snowball state. In addition, daytime temperatures will be higher on dryer planets at the outer HZ regions due to their smaller thermal inertia.

  Combined with the shallow-waters argument, considerations of dry planets thus suggest that planets with a lower fractional surface coverage of water, and with bodies of liquid water that are distributed over many reservoirs rather than combined in one big ocean, can be considered superhabitable.

[5] Note that Earth is located at the very inner margin of the solar habitable zone (Kopparapu et al., 2013).

[6] We here understand superhabitability as a state in which a terrestrial world is generally more habitable than Earth. Conventionally, habitability is considered a binary condition, an "on/off" or "1/0" state, just as a sow is in pig or not. In this sense, we discuss the prospects of a sow being pregnant with several farrows, a state more fertile than only "on" or "1".

- Plate tectonics: On Earth, plate tectonics drive the carbon-silicate cycle. In this planet-wide geochemical reaction, near-surface weathering of calcium silicate ($CaSiO_3$) rocks leads to the formation of quartz-like minerals, that is, silicon dioxide ($SiO_2$). At the same time, carbon dioxide ($CO_2$, for example from the atmosphere) combines with the residual carbon atoms to form calcium carbonate ($CaCO_3$). When subducted to deeper sediments, elevated pressures and temperatures reverse this reaction, ultimately leading to volcanic outgassing of $CO_2$. If this cycle stopped or if it never started on a hypothetical terrestrial, water-rich planet, then silicate weathering would draw down atmospheric $CO_2$, which could lead to a global snowball state. On a planet that receives more stellar illumination or has other internal heat sources (for example tidal or radiogenic heating), this collapse could be avoided. The period over which radiogenic heating is strong enough to maintain plate tectonics increases with increasing planetary mass (Walker et al., 1981). To a certain degree, more massive terrestrial planets should thus tend to be superhabitable.

  However, planets with masses several times that of Earth develop high pressures in their mantle, and the resulting enhanced viscosities make plate tectonics less likely (Noack and Breuer, 2011). Moreover, a stagnant lid forms at the core-mantle-boundary that allows only a reduced heat flow from the core and thereby also frustrates tectonics (Stamenković et al., 2011). Too high a mass thus impedes plate tectonics and therefore also subduction that is required for the carbon-silicate. Yet, "propensity of plate tectonics seems to have a peak between 1 and 5 Earth masses" (Noack and Breuer, 2011), which, of course, depends on composition and primordial heat reservoir. We conclude that planets with masses up to about 2 $M_\oplus$ tend to be superhabitable from the tectonic point of view.

- Magnetic shielding: To allow for surface life, a world must be shielded against high-energy radiation from interstellar space (termed "cosmic radiation") and from the host star (Baumstark-Khan and Facius, 2002). Too strong an irradiation could destroy molecules relevant for life, or it could strip off the world's atmosphere, an effect to which low-mass terrestrial worlds are particularly prone (Luhmann et al., 1992). Protection can be achieved by a global magnetic field, whether it is intrinsic as on Earth or extrinsic as may be the case on moons (Heller and Zuluaga, 2013), and by the atmosphere. While a giant planet's magnetosphere can shield a moon against cosmic rays and stellar radiation, it may itself induce a bombardment of the moon with ionized particles that are trapped in the planet's radiation belt (see Jupiter; Fischer et al., 1996).

  To sustain an intrinsic magnetic field strong enough for protection over billions of years, a terrestrial world needs to have a liquid, rotating, and convecting core. Within Earth, this liquid is composed of molten iron alloys in the outer core, that is, between 800 and 3000 km from its center. Less massive planets or moons will have weaker, short-lived magnetic shields. Williams et al. (1997) estimated a minimum mass of 0.07 $M_\oplus$ for a world under solar irradiation to retain atmospheric oxygen and nitrogen over 4.5 Gyr. Beyond that, the dipole component of the magnetic moment $\mathscr{M}$ depends on the core radius $r_o$, rotation frequency $\Omega$, and the thickness $D$ of the core rotating shell where convection occurs via $\mathscr{M} \propto r_o^3 D^{5/9} \Omega^{7/6}$ (Olson and Christensen, 2006; López-Morales et al., 2011), which implies that tidally locked planets and moons in wide orbits may have weak magnetic shielding.

- Climatic thermostat: A more reliable global thermostat that impedes ice ages and snowball states would prevent an existing ecosystem from experiencing mass extinctions, which would decelerate or even frustrate evolution. There should exist atmospheric and geological processes whose interplay constitutes a thermostat that makes a planet superhabitable.

  Triggered by the recent discoveries of super-Earth planets in or near the stellar HZ, recycling mechanisms of atmospheric $CO_2$ and $CH_4$ have been proposed for potentially water-rich planets (Kaltenegger et al., 2013).[7] These planets are predicted to be completely covered by a deep liquid water ocean on top of high-pressure ices and without direct contact to the rocky interior. On such worlds, an Earth-like carbon-silicate cycle cannot possibly operate as there would be no $CO_2$ weathering. Alternatively, lattices of high-pressure water molecules could trap $CO_2$ as guest molecules, a chemical substance known as carbon clathrate, and provide an effective climatic thermostat by moderating the $H_2O$ and $CO_2$ levels in water-rich super-Earths. A similar clathrate mediation has been shown possible for $CH_4$ instead of $CO_2$ (Levi et al., 2013), that is, methane clathrate. Clathrate convection could be an effective mechanism to transport $CH_4$ and/or $CO_2$ from a water-rich planet's silicate-iron core through a high-pressure ice-mantle into the ocean and, ultimately, into the atmosphere (Fu et al., 2010).

- Surface temperature: On worlds with substantial atmospheres, in other words with surface pressures $P$ at least as high as those on Mars (where 1 mb $\lesssim P \lesssim$ 10 mb), surface temperatures will generally be different from the thermal equilibrium temperature given by stellar irradiation and planetary albedo alone (Selsis et al., 2007; Leconte et al., 2013). The biodiversity, or the richness of families and genera, seems to have multiplied during warmer epochs on Earth (Mayhew et al., 2012), indicating that worlds warmer than Earth could be more habitable. A slightly warmer version of Earth might have extended tropical zones that would allow for more biological variance. This is suggested by both the "cradle model" and the "museum model" used in evolutionary biology. The former approach suggests that rapid diversification occurred recently and rapidly in the tropics, while the latter theory claims that the tropics provide particularly favorable circumstances for slow accumulation and preservation of diversity over time (McKenna and Farrell, 2006; Moreau and Bell, 2013).

[7] As candidates for such water-rich planets, Levi et al. (2013) propose Kepler-11b, Kepler-18, and Kepler-20b. Kaltenegger et al. (2013) suggest Kepler-62e and f.

However, warming Earth does not necessarily yield increased biodiversity. Warming on short timescales causes mass extinction, which can currently be witnessed on Earth. Only a planet that is warm compared to Earth on a Gyr timescale or a world that warms gently over millions and billions of years could have more extended surface regions suitable for liquid water and biodiversity.

On the downside, with fewer temperate zones and no arctic regions, an enormous range of life forms known from Earth could not exist. Above all, a world that is substantially warmer than Earth might have anoxic oceans. On Earth, Oceanic Anoxic Events occurred in periods of warm climate, with average surface temperatures above 25°C compared to pre-industrial 14°C (IPCC, 1995), and resulted in extensive extinctions like the Permian/Triassic around 250 Myr ago (Wignall and Twitchett, 1996). While the concatenation of circumstances that led to extinctions during hot periods is complicated and may reflect problems of Earth's ecosystem, it cannot be excluded that a world moderately warmer than Earth could be superhabitable. A colder planet, however, can be assumed to be less habitable as less energy input would slow down chemical reactions and metabolism on a global scale.

- Biological diversification: An inhabited planet whose flora and fauna are more diverse than they are on Earth could reasonably be termed superhabitable as it empirically shows that its environment is more benign to life. An evolutionary explosion, such as the Cambrian one on Earth, could occur earlier in a planet's history than it did on Earth – or simply long enough ago to make the respective planet more diversely inhabited than Earth is today. Alternatively, evolution could have progressed faster on other planets. Jumps in diversification or accelerated evolution can be triggered by nearby supernovae and by enhanced radiogenic or ultraviolet radiation.

- Multihabitability and panspermia: Stellar systems could be more habitable than the Solar System if there were more than one terrestrial planet or moon in the HZ (Borucki et al., 2013; Anglada-Escudé et al., 2013[8]). If, for example, the Moon-forming impact had distributed the mass more evenly between Earth and the Moon, then both objects might have been habitable. Alternatively, in a hypothetical Solar System analog in which only the orbits of Mars and Venus would be exchanged, there could exist three habitable planets. With the possibility of massive moons about giant planets, there might also exist satellite systems with several habitable exomoons. Such stellar systems could be called "multihabitable." Impacts of comets, asteroids, or other interplanetary debris might trigger exchange of material between those worlds. This exchange could then induce mutual fertilization among multiple habitable worlds, a process known as panspermia (Hoyle and Wickramasinghe, 1981; Weber and Greenberg, 1985). Worlds in multihabitable systems, whether they are planets or moons, could thus be regarded as superhabitable because they have a higher probability to be inhabited.

- Localization in the stellar habitable zone: Recent work emphasized that Earth is scraping at the very inner edge of the Sun's HZ (Kopparapu et al., 2013; Worsworth and Pierrehumbert, 2013). Terrestrial worlds that are located more towards the center of the stellar HZ could be considered superhabitable. These objects would be more resistant against transitioning into a moist or runaway greenhouse state (at the inner edge of the HZ) than Earth is.

- Age: From a biological point of view, older worlds can be assumed to be more habitable because Earth experienced a steady increase in biodiversity as it aged (Mayhew et al., 2012). This diversification indicates that non-intelligent life itself is able to modify an environment so as to make it more suitable for its ancestors.[9] A stronger claim has been put forward by what is now known as the Gaia hypothesis, which suggests that the global biosphere as a whole can be regarded as a creature controlling "the global environment to suit its needs" (Lovelock, 1972). Whether considered as a global entity or not, Earth's ecosystem obviously influences global geochemical processes, which has perpetually led to an increase in biodiversity over billions of years. As an example, note that after the Great Oxygen Event about 2.5 Gyr ago (Anbar et al., 2007)[10], which was likely induced by oceanic algae, Earth's surface became more habitable, allowing life to conquer the continents about 360 to

[8] In the case of GL 667C, it is entertaining to imagine how the structure of that system might influence the development of an intelligent species' astronomy and human space flight activities, with three potentially habitable worlds and three complete stellar systems to study up-close.

[9] Intriguingly, now that the first form of life on Earth able to call itself intelligent it causes a drastic decrease in biodiversity. But even in case evolution typically leads to intelligent life, then if an intelligence destroyed itself, it can be assumed that the respective ecosystem would be able to recover on a Myr or Gyr timescale, of course depending on the magnitude of the caused extinction and the environmental effects left behind by the intelligence.

[10] Analyses of chromium isotopes and redox-sensitive metals of drill cores from South Africa by Crowe et al. (2013) indicate a first slight increase in atmospheric oxygen about 3 Gyr ago, which could be related to the emergence of oxygenic photosynthesis.

480 Myr ago (Kenrick and Crane, 1997). Therefore, older planets should tend to be more habitable, or superhabitable if inhabited.

- Stellar mass: The mass of a star on the main sequence determines its luminosity, its spectral energy distribution, and its lifetime. The Sun emits most of its light between 400 nm and 700 nm, which is the part of the spectrum visible to the human eye. This is also the spectral range in which plants and other organisms perform oxygenic photosynthesis. On worlds orbiting stars with masses $\lesssim 0.6\,M_\odot$ (known as M dwarfs, Baraffe and Chabrier, 1996), these forms of life might not have the capacity to properly harvest energy for their survival because their stars have their radiation maxima in the infrared. However, Miller et al. (2005) found a free-living cyanobacterium that is able to use near-infrared photons at wavelengths > 700 nm. This discovery, as well as the ability of the oxygenic photosynthetic cyanobacterium Acaryochloris marina to use chlorophyll d for harvesting photons at 750 nm (Chen and Blankenship, 2011), suggests that – of course provided that many other conditions are met – oxygenic photosynthesis on planets orbiting cool stars is possible. Discussing the results of Kiang et al. (2007a,b) and Stomp et al. (2007), Raven (2007) also concluded that photosynthesis can occur on exoplanets in the HZ of M dwarfs. Ultimately, the transmissivity of the planet's atmosphere needs to be appropriate to allow an adequate amount of spectral energy to arrive at the planet's surface.

  We will not go deeper in possible extremophile life – extremophile from the standpoint of an Earthling – and, for the time being, consider M stars as less likely hosts for superhabitable planets. However, these reflections show that stars slightly less massive than the Sun could still provide the appropriate spectral energy distribution for photosynthesis.

- Stellar UV irradiation: Stellar UV radiation can damage deoxyribonucleic acid (DNA) and thus impede the emergence of life. Today, Earth has a substantial stratospheric ozone column that absorbs solar irradiation almost completely between 200 and 285 nm (UVC) and most of the radiation between 280 and 315 nm (UVB). During the Archean (3.8 to 2.5 Gyr ago), this ozone shield did not exist, but yet life managed to form. We can assume that terrestrial planets with anoxic primordial atmospheres would be more habitable than early Earth if they received less hazardous UV irradiation.

  M stars remain very active and emit a lot of X-ray and UV radiation during about the first Gyr of their lifetime (Scalo et al., 2007). The activity-driven XUV flux of G stars, such as the Sun, falls off much more rapidly, but their quiescent UV flux is enhanced with respect to K and M dwarfs. What is more, while the UV flux of young M stars is generally much stronger than that of young Sun-like stars, quiescent UV radiation from evolved M dwarfs may be too weak for some essential biochemical compounds to be synthesized (Guo et al., 2010). Thus, they do not seem to offer superhabitable primordial environments. K stars offer a convenient compromise between moderate initial and long-term high-energy radiation. This is supported by considerations of the weighted irradiance spectrum of complex carbon-based molecules, indicating that planets in the HZs of K main sequence stars experience particularly favorable UV environments (Cockell, 1999). This indicates that K dwarf stars are favorable host stars for superhabitable planets.

- Stellar lifetime: With a planet's tendency to be superhabitable increasing with age, the star must burn long enough for existing life forms to evolve. Stars less massive than the Sun have longer lifetimes, and planets or moons can spend more time within the HZ before they transition inside the expanding inner edge (Rushby et al., 2013). Against the background from the two previous items and accounting for the relatively stable spectral radiance once they have settled on the main-sequence, we propose that K dwarfs are more likely to host superhabitable planets than the Sun or M dwarfs.

- Early planetary bombardment: The nature of Earth is closely coupled to its bombardment history. From the lunar forming impact (Cameron and Ward, 1976) to the Late Heavy Bombardment (Gomes et al., 2005), the impact history influenced the surface environment, delivery of organic molecules and volatiles (Chyba and Sagan, 1992; Raymond et al., 2009), and spin/orbital evolution of Earth. This means that the history of Earth's evolution is closely coupled to the orbital dynamics of the planetary system. It is possible that the LHB itself is responsible for Earth's habitability, since it helped deliver water and other volatiles to Earth's surface from farther out in the Solar System.

  While the exact cause of the LHB is uncertain, the debate has focused on effects of a continuous, though gradually tapering, history of impacts versus a spiked delivery of material caused by changes in orbital dynamics (Ryder, 2002). Either way, the system architecture played a large role in determining the extent of these impacts (Raymond et al., 2004). Is it possible that a system with more dynamical instability early in a planet's history would result in a longer, more extensive LHB, or – in the case of a stochastic LHB – a sequence of LHB-type events? Such a history could have little effect on the ongoing evolution of marine or subterranean microbes, yet result in a richer volatile inventory for the host planet or moon, and even encourage multihabitability by enhancing transfer of material between planets in the same system.

- Planetary spin: The initial spin-orbit misalignment, or obliquity, and rotation rate of a planet are largely due to the random events that lead to a planet's formation (Miguel and Brunini, 2010), but the subsequent evolution is tightly coupled to orbital dynamics. Conventional wisdom suggests that Earth is an "ideal" habitable world, since it has a large – and presumably rare

– moon to stabilize its tilt relative against the orbital forcing from the Sun and other planets (Laskar et al., 1993). However, there are a couple of assumptions in this: 1) that a stable spin is required or even desired for a habitable planet, and 2) that this effect is not mitigated by the crucial role the Moon has had on the evolution of Earth's spin rate. For example, studies have indicated that Earth's rotation axis could be stable without the presence of a massive satellite (Lissauer et al., 2012), and that such stability is perhaps not desirable (Spiegel et al., 2009; Armstrong et al., 2013). In the latter case, planets with a large tilt can break the ice-albedo feedback at locations further from the star, keeping the planet from entering the "snowball Earth" stage (Williams and Kasting, 1997), and systems with varying tilts could provide slow but steady changes in ecosystems that encourage evolution of life.

It is uncertain whether any given spin rate is desirable for life, as long as it helps keep the surface uniformly habitable, while radical changes in such a spin rate might be detrimental. Did the existence of the Moon encourage life to evolve by changing the diurnal and tidal cycles, or was this an impediment to evolution? Could moderate changes of a world's obliquity or rotation rate even force life to adapt to a broader range of environmental conditions, thereby triggering more diverse evolution? Ultimately, is it possible that a terrestrial planet without a massive moon, or a planet more subject to changes in spin, could be superhabitable?

- Orbital dynamics: It is occasionally claimed that Earth is habitable largely owing to its stable, circular orbit. However, climate studies indicate a range of dramatic shifts in climate due to subtle changes in Earth's orbit. These oscillations of obliquity, precession, orbital eccentricity, and rotation period – mainly driven by gravitational interaction with the Sun, the Moon, Jupiter, and Saturn – are known as Milankovitch cycles (Hays et al., 1976; Berger, 1976). The very stability of our orbit, in these cases, makes such events treacherous, as Earth may experience only subtle changes again to help rectify the problem. In fact, it is entirely possible that such stability might put the brakes on biological evolution. Planets with eccentric orbits would still provide a range of seasonally viable habitats while perhaps acting as a "vaccine" against life-threatening snowball events. Tidal heating in planets or moons on eccentric orbits may even act as a buffer against transition into a global snowball state (Reynolds et al., 1987; Scharf, 2006; Barnes et al., 2009). Planets with large swings in eccentricity can also influence the planetary tilt, which has its own, perhaps positive, impacts on the habitability of a planet. We thus claim that moderate variations in the orbital elements of a terrestrial world need not necessarily hamper the evolution or inhibit the formation of life. Consequently, we see no terminating argument that Earth's configuration in an almost circular orbit with mild changes in its orbital elements should be considered the most benign situation. Planets that undergo soft variations in their orbital configurations may still be superhabitable.

- Atmosphere: The atmosphere of an exoplanet or exomoon is essential to its surface life, as it serves as a mediator of transport for water and, to a lesser degree, nutrients. Atmospheric composition and the gases' partial pressures will determine surface temperatures and, hence, have a key role in shaping the environment and providing the preconditions for formation and evolution of life.

  Just as an example of how an atmosphere different from that of Earth could make an otherwise similar world superhabitable, note that (i.) enhanced atmospheric oxygen concentration allows a larger range of metabolic networks (Berner et al., 2007); (ii.) variations in the atmospheric oxygen concentration seem to constrain the maximum possible body size of living forms (Harrison et al., 2010; Payne et al., 2011); and (iii.) there are no known multicellular organisms that are strictly anaerobic. Today, Earth's atmosphere contains about 21% oxygen by volume or partial pressure ($pO_2$). Limited by runaway wildfires for $pO_2 > 35\%$ and lack of fire at $pO_2 < 15\%$ (Belcher and McElwain, 2008), a range of oxygen partial pressures is compatible with an ecosystem broadly similar to Earth's. Obviously, atmospheric oxygen contents can be much greater than on Earth today, and worlds with oxygen-rich atmospheres could be entitled superhabitable, because of items (i.) – (iii.).

  While atmospheres less massive than that of Earth would offer weaker shielding against high-energy irradiation from space, weaker balancing of day-night temperature contrasts, retarded global distribution of water, etc., somewhat more massive atmospheres could induce positive effects for habitability. Again, this indicates that planets slightly more massive than Earth should tend to be superhabitable because, first, they acquire thicker atmospheres and, second, their initially extended hydrogen atmospheres can envelope gaseous nitrogen and thereby prevent its loss due to non-thermal ion puck up under an initially strong stellar UV irradiation (Lammer, 2013).

Some of the conditions listed in this Section are already, or will soon be, accessible remotely (namely, orbital and bodily characteristics of extrasolar planets or moons), some will be modeled and thereby constrained (such as orbital evolution and composition), and others will remain hidden and induce random effects on habitability (climate history, radiogenic heating, ocean salinity, former presence of meanwhile ejected planets or satellites, etc.) from the viewpoint of an observer. This list is by far not complete, and it is not our goal to provide such a complete list. However, it is supposed to illustrate that a range of physical characteristics and processes can make a world exhibit more benign environments than Earth does. Given the amount of planets that exist in the Galaxy, it is therefore reasonable to predicate that worlds with more comfortable settings for life than Earth exist.

Earth might still be rare, but this does not make the emergence and existence of extraterrestrial life impossible or even very unlikely because superhabitable worlds exist.

## 4. Conclusions

Utilization of any flavor of the HZ concept implies that a planet is either in the HZ and habitable or outside it and uninhabitable. Resuming our considerations from Section 2, our results are threefold: (i.) Extensions of the HZ concept which include tidal heating, show that planets ("Super-Europas" in our terminology) can exist beyond the HZ and still be habitable. (ii.) Fed by tidal heating, moons of planets beyond the HZ can be habitable. (iii.) Intriguingly, none of all the discussed concepts for the HZ describes a circumstellar distance range that would make a planet a more suitable place for life than Earth currently is.

Terrestrial planets that are slightly more massive than Earth, that is, up to 2 or 3 $M_{\oplus}$, are preferably superhabitable due to the longer tectonic activity, a carbon-silicate cycle that is active on a longer timescale, enhanced magnetic shielding against cosmic and stellar high-energy radiation, their larger surface area, a smoother surface allowing for more shallow seas, their potential to retain atmospheres thicker than that of Earth, and the positive effects of non-intelligent life on a planet's habitability, which can be observed on Earth. Higher biodiversity made Earth more habitable in the long term. If this is a general feature of inhabited planets, that is to say, that planets tend to become more habitable once they are inhabited, a host star slightly less massive than the Sun should be favorable for superhabitability. These so-called K dwarf stars have lifetimes that are longer than the age of the Universe. Consequently, if they are much older than the Sun, then life has had more time to emerge on their potentially habitable planets and moons, and – once occurred – it would have had more time to "tune" its ecosystem to make it even more habitable.

The K1V star Alpha Centauri B (αCenB), member of the closest stellar system to the Sun and supposedly hosting an Earth-mass planet in a 3.235-day orbit (Dumusque et al., 2012), provides an ideal target for searches of planets in the HZ and, ultimately, for superhabitable worlds. Age estimates for αCenB, derived via asteroseismology, chromospheric activity, and gyrochronology (Thévenin et al., 2002; Eggenberger et al., 2004; Thoul et al., 2008; Miglio and Montalbán, 2008; Bazot et al., 2012), show the star to be slightly evolved compared to the Sun, with estimates being 4.85 ±0.5 Gyr, 6.52 ±0.3 Gyr, 6.41 Gyr, 5.2 – 8.9 Gyr, and 5.0 ±0.5 Gyr, respectively. Radiation effects of the stellar primary Alpha Centauri A have been shown to be small and should not induce significant climatic variations on planets about αCenB (Forgan, 2012). If life on a planet or moon in the HZ of αCenB evolved similarly as it did on Earth and if this planet had the chance to collect water from comets and planetesimals beyond the snowline (Wiegert and Holman, 1997; Haghighipour and Raymond, 2007), then primitive forms of life could already have flourished in its waters or on its surface when the proto-Earth collided with a Mars-sized object, thereby forming the Moon.

Eventually, just as the Solar System turned out to be everything but typical for planetary systems, Earth could turn out everything but typical for a habitable or, ultimately, an inhabited world. Our argumentation can be understood as a refutation of the Rare Earth hypothesis. Ward and Brownlee (2000) claimed that the emergence of life required an extremely unlikely interplay of conditions on Earth, and they concluded that complex life would be a very unlikely phenomenon in the Universe. While we agree that the occurrence of another truly Earth-like planet is trivially impossible, we hold that this argument does not constrain the emergence of other inhabited planets. We argue here in the opposite direction and claim that Earth could turn out to be a marginally habitable world. In our view, a variety of processes exists that can make environmental conditions on a planet or moon more benign to life than is the case on Earth.

## Appendix A: Usage and Meaning of Terms Related to Habitability

Discussions about habitability suffer from diverging understanding of the terms "habitability," "habitable," etc. Recall that a planet in the stellar illumination HZ, as it is defined by physicists and astronomers (see Section 2), need not necessarily be habitable. It is thus precipitate, if not simply false, to state that the planet Gl581d is "habitable, but not much like home" (Schilling, 2007). Analogously, a world such as a tidally heated moon outside the HZ need not necessarily be uninhabitable. Claiming that "Being inside the habitable zone is a necessary but not sufficient condition for habitability" (Selsis et al., 2007) can be wrong, depending on the meaning of the word "habitable." If that statement means that habitable planets are in the HZ by definition, then the sentence is tautological. If, however, it means that a planet needs to be in the HZ to provide liquid surface water, then it can be proven wrong.

Confusions from blurred pictures are not restricted to the qualitative. As an example, quantitative problems occur in discussions about the occurrence rate of planets similar to Earth that orbit Sun-like stars. The parameter $\eta_{\oplus}$ has been introduced to quantify their abundance. Unfortunately, different understandings of $\eta_{\oplus}$ occur in the literature. It has been used as "fraction of stars with Earth-mass planets in the habitable zone" (Howard et al., 2009), "the fraction of Sun-like stars that have planets like Earth" (Catanzarite and Shao, 2011), "the fraction of Sun-like stars with Earth-like planets in their habitable zones" (O'Malley-James et al., 2012), "the fraction of habitable planets for all Sun-like stars" (Catanzarite and Shao, 2011), "the fraction of Sun-like stars that have at least one planet in the habitable zone" (Lunine et al., 2008), the "frequency of Earth-mass planets in the habitable zone" (Wittenmyer et al., 2011), the fraction of "Earth-like planets with $M \sin i$ = 0.5-2$M_{Earth}$ and $P < 50$ days"[11] (Howard et al., 2010), "the frequency of habitable planets orbiting M dwarfs" (Bonfils et al., 2013b), "the frequency of $1 < m \sin i < 10\,M_{\oplus}$ planets in the habitable zone of M dwarfs"[12] (Bonfils et al., 2013a), "the frequency of terrestrial planets in the habitable zone [...] of solar-like stars in our galaxy" (Jenkins, 2012), and "the number of planets with $0.1M_{\oplus} < M_p < 10M_{\oplus}$ in the 3 Gyr CHZ (a < 0.02AU)"[13] (Agol, 2011). The latter two definitions stand out because Jenkins (2012) restricts $\eta_{\oplus}$ to the Milky Way, and Agol (2011) introduces $\eta_{\oplus}$ as a total count, and yet, he uses it as a frequency.

Intriguingly, (i.) as it is not clear whether a planet must be similar to Earth to be habitable, (ii.) as the definitions diverge in their reference to the stellar type, and (iii.) as it sometimes remains obscure what "Earth-like planets" are in the respective context, none of these understandings is equivalent to at least one of the others, except for the Howard et al. (2009) and Wittenmyer et al. (2011) explanations. As a consequence, different estimates for $\eta_{\oplus}$ *must* occur. Although physical, observational, and systematic effects play a role, a quantitative divergence of estimates for $\eta_{\oplus}$ will remain as long as there is no consensus about the meaning, that is, the usage, of this word or variable. This problem is not physical, but it is a logical consequence of the diverging understanding of $\eta_{\oplus}$. Imagine a situation in which all the authors of the mentioned studies sit around a desk to discuss their values for $\eta_{\oplus}$ and the implications! If they were not aware of the meaning/usage drift of "their" respective $\eta_{\oplus}$, then their dialogue would founder on a language problem.

The crux of the matter lies in the meaning of any of these terms, which again depends on the context in which any term is used. Following the Austrian philosopher Ludwig Wittgenstein and his *Philosophische Untersuchungen* (Wittgenstein, 1953), many logical problems occur when terms are alienated from their ancestral use and then unreasonably applied in other contexts. Ultimately, as astrobiology is an interdisciplinary science, it is exposed to those dangers of confusion and contradiction to a special degree. In this communication, we shall not infringe the use of language and terminology but unravel possible perils. In other words, we ought to be descriptive rather than normative (Wittgenstein, 1953, §124). To answer the question of whether a planet is habitable, it must be clear what we understand a habitable planet to be. And following semantic holism, a doctrine in the philosophy of language, the term "habitable" then is defined by its usage in the language.[14]

## Appendix B: An Algebraic Approach to Superhabitable Planets

Astronomers have developed an inclination to evaluate habitability in terms of geocentric conditions. Expressions such as "Earth-like," "Earth analog," "Earth twin," "Earth-sized," and "Earth-mass" are often used to evaluate a planet's habitability. Although being a natural body of reference, if other inhabited worlds exist – and obviously some scientists assume that and look for them – then it would be presumptuous to claim that they need to be Earth-like or that Earth offers the most favorable conditions. We can use set algebra to discern and display planet families. This somewhat unconventional approach would allow us to identify Earth as one sort of an habitable and inhabited world and to become acquainted with superhabitable worlds.

### *Appendix B.1: Set theory*

Consider a set T of terrestrial planets. We assume that any solar or extrasolar planet will either be an element of T or not. Planets have been detected with masses of about 5 to 10 Earth masses, and they likely constitute a transitional regime between terrestrial and, as the case may be, icy or gaseous. They may still have their bulk mass in solid form but also have a substantial atmosphere. Nevertheless, we use a sharp classification here for simplicity. We concentrate here on the genuine terrestrial planets. As an example, Earth ($e_{\oplus}$) is a terrestrial planet ($e_{\oplus} \in T$), whereas Jupiter is not.

The elements of T are the terrestrial planets: $T = \{t \in T \mid t \text{ terrestrial}\}$ (see Fig. A). Some of these planets will be habitable and thus be an element of the set of habitable, terrestrial planets $H = \{h \in T \mid h \text{ habitable}\}$ (dotted area). The complement of this set is the set of uninhabitable, terrestrial planets $U = \bar{H} = \{u \in T \mid u \text{ uninhabitable}\}$ (blank area). There are no planets that are both habitable and uninhabitable. Hence, the union of H and U is equal to the terrestrial planets: $H \cap U = T$. Beyond, there will be a set of Earth-like planets $E = \{e \in T \mid e \text{ Earth-like}\}$ (vertically striped area). Our intuition, trained by the usage of the term "Earth-like" in literature, by talks, and conversations, suggests that Earth-like planets are habitable. For the time being, we prefer to take a more general point of view and allow Earth-like planets also to be uninhabitable. E thus overlaps with U in Fig.

[11] In this context, M is planetary mass, i is the inclination of the planet's orbital plane with respect to an Earth-based observer's line of sight, $M_{Earth}$ is an Earth-mass, and *P* is the planet's orbital period about the star.

[12] Here, m is planetary mass and *i* the inclination of the planet's orbital plane with respect to an Earth-based observer's line of sight.

[13] In this context, $M_p$ is planetary mass, "CHZ" is an abbreviation for the "continuous habitable zone", and a is the planet's orbital semi-major axis.

[14] (Wittgenstein, 1953, §43): "Die Bedeutung eines Wortes ist sein Gebrauch in der Sprache."

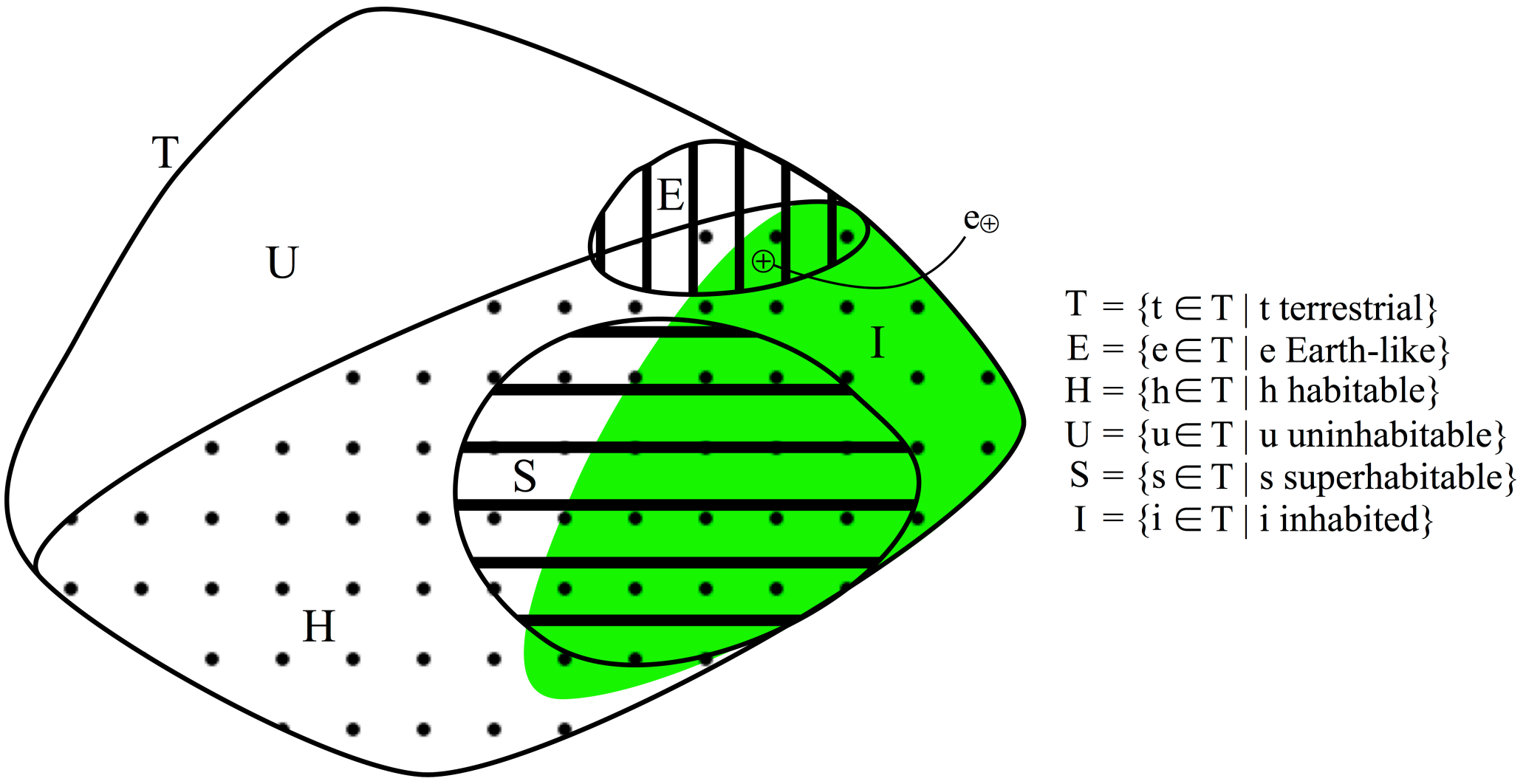

**Fig. A:** Set of terrestrial worlds T and subsets. The set membership of Earth $e_{\oplus} \in (E \cap I)$ is indicated with a symbol. This graphic visualizes our claim that habitable planets (H) need not be Earth-like (E), and that there may well exist a set of superhabitable worlds (S). The cardinality of S may be greater than that of E, and the fraction of planets inside S that are actually inhabited (I, green) may be greater than the fraction of Earth-like, inhabited planets. For this purpose, S is depicted to be larger than E, and $(E \cap I)$ is chosen to be smaller with respect to E than $(S \cap I)$ with respect to S.

A. Yet, to be inhabited, a terrestrial planet must also be habitable. Thus, the set $I = \{i \in T \mid i \text{ inhabited}\}$ (green area) of inhabited planets is a subset of H: $I \subseteq H$. Note that the equality is only valid if all the habitable planets were indeed inhabited. It is reasonable to assume that there exists at least one terrestrial planet that is habitable but yet uninhabited. Thus, we can securely state $I \subset H \Leftrightarrow (t \in I \Rightarrow t \in H)$.[15] With Earth being Earth-like, habitable, and inhabited, we have $e_{\oplus} \in (H \cap I \cap E)$. Finally, we propose that there exists a set $S = \{s \in T \mid s \text{ superhabitable}\}$ (horizontally striped area) of terrestrial planets, whose elements (that is, superhabitable planets) offer more comfortable environments to life than Earth does. From a statistical perspective, this statement reads:

A randomly chosen element $s \in S$ is more likely to be inhabited than a randomly chosen element $e \in E$. (1)

Alternatively, with $p$ being the probability of a planet to be inhabited:

$$p(s) > p(e) \qquad (s \in S,\ e \in E) \tag{2}$$

In Fig. A, we insinuate sentences (1) and (2) by plotting the relative area of $(S \cap I)$ to S larger than the relation of $(E \cap I)$ to E. An equivalent sentence to (2) is

$$|S \cap I| \,/\, |S| = p(s) > |E \cap I| \,/\, |E| = p(e) \qquad (s \in S,\ e \in E), \tag{3}$$

where $|X|$ is the number of elements, or "cardinality," of X.

Sentences (1) – (3) say nothing about the absolute number of inhabited worlds from sets E and S, which corresponds to the size of the areas of E and S in Fig. A. Perhaps there are only two superhabitable planets in our galactic neighborhood, both of which are inhabited, and it may be that there are one hundred Earth-like planets in a similar volume, of which, say, ten are inhabited. Then still (2) is true because $p(s) = 2/2 = 1 > p(e) = 10/100 = 0.1$. But there would be five times as many Earth-like planets with life than there are superhabitable inhabited planets.

In debates about habitable planets, it is subliminally assumed that there are more Earth-like inhabited planets than there are non-Earth-like inhabited planets: $|E \cap I| > |\bar{E} \cap I|$. However, the numbers $|E \cap I|$, $|E \cap \bar{I}|$, $|(\bar{E} \cap T) \cap I|$, and $|(\bar{E} \cap T) \cap \bar{I}|$ are truly not known, say for a local volume of 100 pc about the Sun. There are only the following constraints: $|E \cap I| \geq 1$ and $|(\bar{E} \cap T) \cap \bar{I}| \geq 30 =$ |{CoRoT-7b, Kepler-10b, 55Cnc e, Kepler-18b, Kepler-20e, Kepler-20f, Kepler-36b, Kepler-42b, Kepler-42c, Kepler-42d,

[15] This question of equality is related to the question how long it took life to occur on Earth after the planet became habitable. In fact, planets may generally become inhabited very shortly after becoming habitable. This would allow one to advocate the $I \subseteq H$ relation or even $I = H$.

Kepler-62c, and others}|[16] (Léger et al., 2009; Batalha et al., 2011; Winn et al., 2011; Cochran et al., 2011; Fressin et al., 2012; Carter et al., 2012; Muirhead et al., 2012; Borucki et al., 2013). More terrestrial planet candidates are known, but they lack either radius or mass determinations (for example Gl581d, GJ667Cb to h, GJ1214b, HD 88512, and Alpha Centauri Bb).

The possible existence of S has fundamental observational implications. Were it possible to describe S and predict the characteristics of its elements s, as we attempt in this communication, then the search for extraterrestrial life could be made more efficient. Assume two planets were found; one ($\hat{e}$) being Earth-like and another one ($\hat{s}$) being member of S. Then it would be more reasonable to spend research resources on $\hat{s}$ rather than on $\hat{e}$ in order to find extrasolar life. And intriguingly, $\hat{s}$ could be less Earth-like than $\hat{e}$. Ultimately, a superhabitable world may already have been detected but not yet noticed as such.

*Appendix B.2: The principle of mediocracy*

The principle of mediocracy claims that, if an item is drawn at random from one of several categories, it is likelier to come from the most numerous category than from any of the other less numerous categories (Section 1 in Kukla, 2010). As an example, consider the cardinalities of two sets A and B were known; $|A| < |B|$ and $A \cap B = \varnothing$, where $\varnothing$ is the empty set. Further, $A \cup B = M = \{m \mid m \in A \vee m \in B\}$ is the set of all elements. Then if an arbitrary element $\acute{m} \in M$ were drawn, it would be more likely to come from B than from A. This is all the principle of mediocracy states. In this reading, it comes as a truism. Note that the proportion of A and B, that is, the prior $|A| < |B|$, is known and it is the probability for the drawing that is inferred: $p(\acute{m} \in B) > p(\acute{m} \in A)$.

In a second reading of the principle of mediocracy, and this is the one subliminally applied in modern searches for inhabited planets, the functions of the prior and the drawing are reversed. Here, $\acute{m}$ (which in our example from Section B.1 is Earth, $e_\oplus$) has already been drawn. It is recognized as an element of a certain set $(E \cap I)$, and it is claimed that this set is more abundant than the other one. In the terrestrial worlds scenario (Section B.1), this ventured conclusion reads "$e_\oplus \in (E \cap I) \Rightarrow |E \cap I| > |\bar{E} \cap I|$". We paraphrase it because it is not justified. Given that we have almost no antecedent knowledge of E, $\bar{E}$ , and I, this claim is not logic.[17]

What is more, it is not logical to state that the choice of $e_\oplus$ has been random; humans have not chosen the Earth by random from a set T (Mash, 1993). To make things worse, even if we could have chosen $e_\oplus$ randomly from T and if our assumption were correct, then what could we conclude from only one drawing? Numerous drawings, in other words observations and knowledge about inhabitance of many Earth-like and non-Earth-like planets, would be required to reconstruct the prior with statistical significance. Hence, Earth cannot be justified as a reference for astrobiological investigations with the principle of mediocracy. The claim "$e_\oplus \in (E \cap I) \Rightarrow |E \cap I| > |\bar{E} \cap I|$" remains arbitrary and current searches for life might not be designed optimally.

To conclude, the principle of mediocracy cannot explain why Earth should be considered a particularly benign, inhabited world. When applied to our set of terrestrial worlds, the principle simply states that a randomly chosen world most likely comes from the most numerous subset of worlds. In this understanding, the cardinality of the subsets of terrestrial worlds is the prior – it is know before the drawing – and the probability of affiliation with any subset can be predicted. Yet, concluding that inhabited worlds are most likely Earth-like is not logical, because, first, the roles of the prior (here: the inhabited worlds) and the drawing (here: Earth) are reversed and, second, Earth has not been drawn (by whom?) at random.

## Acknowledgments

René Heller thanks Morten Mosgaard for board and lodge on the Danish island Langeland where this study has been initiated. René Heller is funded by the Canadian Astrobiology Training Program and a member of the Origins Institute at McMaster University. Discussions with Rory Barnes have been a valuable stimulation to this study and we appreciate Alyssa Cobb's helpful comments on the manuscript. Computations have been performed with ipython 0.13 (Pérez and Granger, 2007) on python 2.7.2 and figures have been prepared with gnuplot 4.6 (www.gnuplot.info). This work has made use of NASA's Astrophysics Data System Bibliographic Services. Our collaboration has been inspired by a question of John Armstrong asked online during an AbGradCon talk 2012.

## References

Abe, Y., Abe-Ouchi, A., Sleep, N.H., and Zahnle, K.J. (2011) Habitable Zone Limits for Dry Planets. *Astrobiology* 11:443-460. doi:10.1089/ast.2010.0545

Agol, E. (2011) Transit Surveys for Earths in the Habitable Zones of White Dwarfs. *Astrophysical Journal* 731:L31. doi:10.1088/2041-8205/731/2/L31

[16] www.exoplanet.eu/catalog as of July 15, 2013

[17] Yet, it could be true.

Anbar, A.D., Duan, Y., Lyons, T.W., et al. (2007) A Whiff of Oxygen Before the Great Oxidation Event? *Science* 317:1903-1906. doi:10.1126/science.1140325

Anglada-Escudé, G., Tuomi, M., Gerlach, E., et al. (2013) A dynamically-packed planetary system around GJ 667C with three super-Earths in its habitable zone. *Astronomy & Astrophysics* 556:A126 (24 pp.). doi:10.1051/0004-6361/201321331

Armstrong, J.C., Domagal-Goldman, S., Barnes, R., Quinn, T.R., and Meadows, V.S. (2013) Tilt-a-Worlds: Effects of Extreme Obliquity Change on the Habitability of Extrasolar Planets (in preparation)

Baraffe, I. and Chabrier, G. (1996) Mass-Spectral Class Relationship for M Dwarfs. *Astrophysical Journal Letters* 461:L51. doi:10.1086/309988

Barnes, R., Raymond, S.N., Jackson, B., and Greenberg, R. (2008) Tides and the Evolution of Planetary Habitability. *Astrobiology* 8:557-568. doi:10.1089/ast.2007.0204

Barnes, R., Jackson, B., Greenberg, R., and Raymond, S.N. (2009) Tidal limits to planetary habitability. *Astrophysical Journal* 700:L30-L33. doi:10.1088/0004-637X/700/1/L30

Barnes, R., Mullins, K., Goldblatt, C., et al. (2013) Tidal Venuses: Triggering a Climate Catastrophe via Tidal Heating. *Astrobiology* 13:225-250. doi: 10.1089/ast.2012.0851

Batalha, N.M., Borucki, W.J., Bryson, S.T., et al. (2011) Kepler's First Rocky Planet: Kepler-10b. *Astrophysical Journal* 729:27 (21pp.). doi:10.1088/0004-637X/729/1/27

Baumstark-Khan, C. and Facius, R. (2002) Life under conditions of ionizing radiation. In *Astrobiology. The quest for the conditions of life*, edited by Gerda Horneck and Christa Baumstark-Khan, Springer, Berlin, ADS:2002abqc.book..261B

Bazot, M., Bourguignon, S., and Christensen-Dalsgaard, J. (2012) A Bayesian approach to the modelling of α Cen A. *Monthly Notices of the Royal Astronomical Society* 427:1847-1866. doi:10.1111/j.1365-2966.2012.21818.x

Belcher, C.M. and McElwain, J.C. (2008) Limits for Combustion in Low $O_2$ Redefine Paleoatmospheric Predictions for the Mesozoic. *Science* 321:1197-1200. doi:10.1126/science.1160978

Berger, A.L. (1976) Obliquity and precession for the last 5 000 000 years. *Astronomy & Astrophysics* 51:127-135. ADS:1976A&A....51..127B

Berner, R.A., VandenBrooks, J.M., and Ward, P.D. (2007) Oxygen and Evolution. *Science* 316:557-558. doi:10.1126/science.1140273

Bonfils, X., Bouchy, F., Delfosse, X., et al. (2013a) Prized results from HARPS. *EPJ Web of Conferences* 47:05004 (4 pp.). doi:10.1051/epjconf/20134705004

Bonfils, X., Delfosse, X., Udry, S., et al. (2013b) The HARPS search for southern extra-solar planets XXXI. The M-dwarf sample. *Astronomy & Astrophysics* 549:A109. doi:10.1051/0004-6361/201014704

Borucki, W.J., Agol, E., Fressin, F., et al. (2013) Kepler-62: A Five-Planet System with Planets of 1.4 and 1.6 Earth Radii in the Habitable Zone. *Science* 340:587-590. doi:10.1126/science.1234702

Brasier, M., McLoughlin, N., Green, O., and Wacey D (2006) A fresh look at the fossil evidence for early Archaean cellular life. *Philosophical Transactions of the Royal Society B* 361:887-902. doi:10.1098/rstb.2006.1835

Carter, J.A., Agol, E., Chaplin, W.J., et al. (2012) Kepler-36: A Pair of Planets with Neighboring Orbits and Dissimilar Densities. *Science* 337:556-559. doi:10.1126/science.1223269

Cameron, A.G.W. and Ward, W.R. (1976) The Origin of the Moon. *Abstracts of the Lunar and Planetary Science Conference* 7:120-122. ADS:1976LPI.....7..120C

Cassidy, T.A., Mendez, R., Phil, A., Johnson, R.E., and Skrutskie, M.F. (2009) Massive Satellites of Close-In Gas Giant Exoplanets. *Astrophysical Journal* 704:1341-1348. doi:10.1088/0004-637X/704/2/1341

Catanzarite, J. and Shao, M. (2011) The Occurrence Rate of Earth Analog Planets Orbiting Sun-like Stars. *Astrophysical Journal* 738:151. doi:10.1088/0004-637X/738/2/151

Chen, M. and Blankenship, R.E. (2011) Expanding the solar spectrum used by photosynthesis. *Trends in Plant Science* 16:427-431. doi:10.1016/j.tplants.2011.03.011

Chyba, C. and Sagan, C. (1992) Endogenous production, exogenous delivery and impact-shock synthesis of organic molecules: an inventory for the origins of life. *Nature* 355:125-132. doi:10.1038/355125a0

Cochran, W.D., Fabrycky, D.C., Torres, G., et al. (2011) Kepler-18b, c, and d: A System of Three Planets Confirmed by Transit Timing Variations, Light Curve Validation, Warm-Spitzer Photometry, and Radial Velocity Measurements. *Astrophysical Journal Supplements* 197:7 (19 pp.) doi:10.1088/0067-0049/197/1/7

Cockell, C.S. (1999) Carbon Biochemistry and the Ultraviolet Radiation Environments of F, G, and K Main Sequence Stars. *Icarus* 141:399-407. doi:10.1006/icar.1999.6167

Crowe, S.A., Døssing, L.N, Beukes, N.J., et al. (2013) Atmospheric oxygenation three billion years ago. *Nature* 501:535-538. doi:10.1038/nature12426

Des Marais, D.J., Harwit, M.O., Jucks, K.W, et al. (2002) Remote Sensing of Planetary Properties and Biosignatures on Extrasolar Terrestrial Planets. *Astrobiology* 2:153-181. doi:10.1089/15311070260192246

Dole, S.H. (1964) Habitable planets for man. Blaisdell Pub. Co. [1st ed.], New York. ISBN:978-0-8330-4227-9

Dartnell, L. (2011) Biological constraints on habitability. *Astronomy & Geophysics* 52:1.25-1.28. doi:10.1111/j.1468-4004.2011.

52125.x
Debes, J.H. and Sigurdsson, S. (2007) The Survival Rate of Ejected Terrestrial Planets with Moons. *Astrophysical Journal* 668:L167-L170. doi:10.1086/523103
Dressing, C.D., Spiegel, D.S., Scharf, C.A., Menou, K., and Raymond, S.S. (2010) Habitable Climates: The Influence of Eccentricity. *Astrophysical Journal* 721:1295. doi:10.1088/0004-637X/721/2/1295
Dumusque, X., Pepe, F., Lovis, C., et al. (2012) An Earth-mass planet orbiting α Centauri B. *Nature* 491:207-211 doi:10.1038/nature11572
Eggenberger, P., Charbonnel, C., Talon, S., et al. (2004) Analysis of α Centauri AB including seismic constraints. *Astronomy & Astrophysics* 417:235-246. doi:10.1051/0004-6361:20034203
Fischer, H.M., Pehlke, E., Wibberenz, G., Lanzerotti, L.J., and Mihalov, J. (1996) High-Energy Charged Particles in the Innermost Jovian Magnetosphere. *Science* 272:856-858. doi:10.1126/science.272.5263.856
Forgan, D. (2012) Oscillations in the habitable zone around α Centauri B. *Monthly Notices of the Royal Astronomical Society* 422:1241-1249 doi:10.1111/j.1365-2966.2012.20698.x
Fortney, J.J., Marley, M.S., and Barnes, J.W. (2007) Planetary Radii across Five Orders of Magnitude in Mass and Stellar Insolation: Application to Transits. *Astrophysical Journal* 659:1661-1672. doi:10.1086/512120
Fressin, F., Torres, G., Rowe, J.F., et al. (2012) Two Earth-sized planets orbiting Kepler-20. *Nature* 482:195-198. doi:10.1038/nature10780
Fu, R., O'Connell, R.J., and Sasselov, D.D. (2010) The Interior Dynamics of Water Planets. *Astrophysical Journal* 708:1326-1334. doi:10.1088/0004-637X/708/2/1326
Gaidos, E., Deschenes, B., Dundon, L., et al. (2005) Beyond the Principle of Plentitude: A Review of Terrestrial Planet Habitability. *Astrobiology* 5:100-126. doi:10.1089/ast.2005.5.100
Goldblatt, C., Watson, A.J. (2012) The runaway greenhouse: implications for future climate change, geoengineering and planetary atmospheres. *Philosophical Transactions of the Royal Society A* 370:4197-4216. doi:10.1098/rsta.2012.0004
Gomes, R., Levison, H.F., Tsiganis, K., and Morbidelli, A. (2005) Origin of the cataclysmic Late Heavy Bombardment period of the terrestrial planets. *Nature* 435:466-469. doi:10.1038/nature03676
Gray, J.S. (1997) Marine biodiversity: patterns, threats, and conservation needs. *Biodiversity & Conservation* 6:153-175. doi:10.1023/A:1018335901847
Guo, J., Zhang, F., Zhang, X., and Han, Z. (2010) Habitable zones and UV habitable zones around host stars. *Astrophysics and Space Science* 325:25-30. doi:10.1007/s10509-009-0173-9
Haghighipour, N. and Raymond, S.N. (2007) Habitable Planet Formation in Binary Planetary Systems. *Astrophysical Journal* 666:436-446. doi:10.1086/520501
Harrison, J.F., Kaiser, A., and VandenBrooks, J.M. (2010) Atmospheric oxygen level and the evolution of insect body size. *Proceedings of The Royal Society B* 277:1937-1946. doi:10.1098/rspb.2010.0001
Hays, J.D., Imbrie, J., and Shackleton, N.J. (1976) Variations in the Earth's Orbit: Pacemaker of the Ice Ages. *Science* 194:1121-1132. doi:10.1126/science.194.4270.1121
Heller, R., Leconte, J., and Barnes, R. (2011) Tidal obliquity evolution of potentially habitable planets. *Astronomy & Astrophysics* 528:A27:1-16. doi:10.1051/0004-6361/201015809
Heller, R. (2012) Exomoon habitability constrained by energy flux and orbital stability. *Astronomy & Astrophysics* 545:L8. doi:10.1051/0004-6361/201220003
Heller, R. and Barnes, R. (2013a) Exomoon habitability constrained by illumination and tidal heating. *Astrobiology* 13:18-46. doi:10.1089/ast.2012.0859
Heller, R. and Zuluaga, J. (2013) Magnetic Shielding of Exomoons beyond the Circumplanetary Habitable Edge. *Astrophysical Journal Letters* 776:L33 (6 pp.). doi:10.1088/2041-8205/776/2/L33
Heller, R. and Barnes, R. (2013b) Runaway greenhouse effect on exomoons due to irradiation from hot, young giant planets. *International Journal of Astrobiology* (in press). arXiv:1311.0292
Henning, W.G., O'Connell, R.J., and Sasselov, D.D. (2009) Tidally Heated Terrestrial Exoplanets: Viscoelastic Response Models. *Astrophysical Journal* 707:1000-1015. doi:10.1088/0004-637X/707/2/1000
Hinkel, N.R. and Kane, S.R. (2013) Habitability of Exomoons at the Hill or Tidal Locking Radius. *Astrophysical Journal* 774:27 (10 pp.). doi:10.1088/0004-637X/774/1/27
Howard, A.W., Johnson, J.A., Marcy, G.W., et al. (2009) The NASA-UC Eta-Earth Program. I. A Super-Earth Orbiting HD 7924. *Astrophysical Journal* 696:75-83. doi:10.1088/0004-637X/696/1/75
Howard, A.W., Marcy, G.W., Johnson, J.A., et al. (2010) The Occurrence and Mass Distribution of Close-in Super-Earths, Neptunes, and Jupiters. *Science* 330:653-655. doi:10.1126/science.1194854
Hoyle, F. and Wickramasinghe, C. (1981) Space travellors, the bringers of life. Cardiff (UK), University College Cardiff Press. ISBN: 9780906449271. ADS:1981stbl.book.....H
Huang, S.S. (1959) Occurrence of Life in the Universe. *American Scientist* 47:397-402
Huang, S.S. (1960) The Sizes of Habitable Planets. *Publications of the Astronomical Society of the Pacific* 72:489-493. ADS:

1960PASP...72..489H
IPCC (1995) Projected Changes in Global Temperature, Second Assessment Report. www.grida.no/publications/vg/climate
Jackson, B., Barnes, R., and Greenberg, R. (2008a) Tidal heating of terrestrial extrasolar planets and implications for their habitability. *Monthly Notices of the Royal Astronomical Society* 391:237-245. doi:10.1111/j.1365-2966.2008.13868.x
Jackson, B., Greenberg, R., and Barnes, R. (2008b) Tidal heating of Extrasolar Planets. *Astrophysical Journal* 681:1631-1638. doi:10.1086/587641
Jenkins, J.M. (2012) A Waypoint on the Road to Eta$_{Earth}$: Improving the Sensitivity of Kepler's Science Pipeline. AAS Meeting #220, #318.05. ADS:2012AAS...22031805J
Joshi, M.M. and Haberle, R.M. (2012) Suppression of the Water Ice and Snow Albedo Feedback on Planets Orbiting Red Dwarf Stars and the Subsequent Widening of the Habitable Zone. *Astrobiology* 12:3-8. doi:10.1089/ast.2011.0668
Kaltenegger, L. (2010) Characterizing Habitable Exomoons. *Astrophysical Journal Letters* 712:L125-L130. doi:10.1088/2041-8205/712/2/L125
Kaltenegger, L., Selsis, F., Fridlund, M., et al. (2010) Deciphering Spectral Fingerprints of Habitable Exoplanets. *Astrobiology* 10:89-102. doi:10.1089/ast.2009.0381
Kaltenegger, L. and Traub, W.A. (2009) Transits of Earth-like Planets. *Astrophysical Journal* 698:519-527. doi:10.1088/0004-637X/698/1/519
Kaltenegger, L., Sasselov, D., Rugheimer, S. (2013) Water Planets in the Habitable Zone: Atmospheric Chemistry, Observable Features, and the Case of Kepler-62e and -62f. *Astrophysical Journal Letters* 775:L47 (5 pp.). doi:10.1088/2041-8205/775/2/L47
Kasting, J.F., Whitmire, D.P., and Reynolds, R.T. (1993) Habitable Zones around Main Sequence Stars. *Icarus* 101:108-128. doi:10.1006/icar.1993.1010
Kawahara, H., Matsuo, T., Takami, M., et al. (2012) Can Ground-based Telescopes Detect the Oxygen 1.27 μm Absorption Feature as a Biomarker in Exoplanets? *Astrophysical Journal* 758:13. doi:10.1088/0004-637X/758/1/13
Kenrick, P. and Crane, P.R. (1997) The origin and early evolution of plants on land. *Nature* 389:33-39. doi:10.1038/37918
Kiang, N.Y., Siefert, J., Govindjee, and Blankenship, R.E. (2007a) Spectral Signatures of Photosynthesis. I. Review of Earth Organisms. *Astrobiology* 7:222-251. doi:10.1089/ast.2006.0105
Kiang, N.Y., Segura, A., Tinetti, G., et al. (2007b) Spectral Signatures of Photosynthesis. II. Coevolution with Other Stars And The Atmosphere on Extrasolar Worlds. *Astrobiology* 7:252-274. doi:10.1089/ast.2006.0108
Kopparapu, R.K., Ramirez, R., Kasting, J., et al. (2013) Habitable Zones Around Main-Sequence Stars: New Estimates. *Astrophysical Journal* 765:131 (16 pp.). doi:10.1088/0004-637X/765/2/131
Kukla, A. (2010) Extraterrestrials: A Philosophical Perspective. Lexington Books, Rowman & Littlefield Publishers, Inc., Lanham, MD USA. ISBN:978-0-7391-4244-8. ADS:2010epp..book.....K
Lammer, H., Bredehöft, J.H., Coustenis, A., et al. (2009) What makes a planet habitable? *The Astronomy & Astrophysics Review* 17:181-249. doi:10.1007/s00159-009-0019-z
Lammer, H. (2013) Origin and Evolution of Planetary Atmospheres: Implications for Habitability. SpringerBriefs in Astronomy, ISBN 978-3-642-32086-6. doi:10.1007/978-3-642-32087-3
Laskar, J., Joutel, F., Robutel, P. (1993) Stabilization of the earth's obliquity by the moon. *Nature* 361:615-617. doi:10.1038/361615a0
Leconte, J., Chabrier, G., Baraffe, I., and Levrard, B. (2010) Is tidal heating sufficient to explain bloated exoplanets? Consistent calculations accounting for finite initial eccentricity. *Astronomy & Astrophysics* 516:A64. doi:10.1051/0004-6361/201014337
Leconte, J., Forget, F., Charnay, B., et al. (2013) 3D climate modeling of close-in land planets: Circulation patterns, climate moist bistability, and habitability. *Astronomy & Astrophysics* 554:A69 (17 pp.). doi:10.1051/0004-6361/201321042
Léger, A., Rouan, D., Schneider, J, et al. (2009) Transiting exoplanets from the CoRoT space mission. VIII. CoRoT-7b: the first super-Earth with measured radius. *Astronomy & Astrophysics* 506:287-302. doi:10.1051/0004-6361/200911933
Levi, A., Sasselov, D., Podolak, M. (2013) Volatile Transport inside Super-Earths by Entrapment in the Water-ice Matrix. *Astrophysical Journal* 769:29 (9 pp.). doi:10.1088/0004-637X/769/1/29
Lissauer, J.J., Barnes, J.W., Chambers, J.E. (2012) Obliquity variations of a moonless Earth. *Icarus* 217:77-87. doi:10.1016/j.icarus.2011.10.013
López-Morales, M., Gómez-Pérez, N., Ruedas, T. (2011) Magnetic fields in Earth-like Exoplanets and Implications for Habitability around M-dwarfs. *Origins of Life and Evolution of Biospheres* 41:533-537. doi:10.1007/s11084-012-9263-8
Lovelock, J.E. (1972) Gaia as seen through the atmosphere. *Atmospheric Environment Pergamon Press* 6:579-580. doi:10.1016/0004-6981(72)90076-5
Luhmann, J.G., Johnson, R.E., and Zhang, M.H.G. (1992) Evolutionary impact of sputtering of the Martian atmosphere by O(+) pickup ions. *Geophysical Research Letters* 19:2151-2154. doi:10.1029/92GL02485
Lunine, J.I., Fischer, D., Hammel, H.B., et al. (2008) Worlds Beyond: A Strategy for the Detection and Characterization of Exoplanets Executive Summary of a Report of the ExoPlanet Task Force Astronomy and Astrophysics Advisory Committee Washington, DC June 23, 2008. *Astrobiology* 8:875-881. doi:10.1089/ast.2008.0276
Mash, R. (1993) Big numbers and induction in the case for extraterrestrial intelligence. *Philosophy of Science* 60:204-222.
Mayhew, P.J., Bell, M.A., Benton, T.G., McGowan, A.J. (2012) Biodiversity tracks temperature over time. *Proceedings of the*

*National Academy of Sciences* 109:15141-15145. doi:10.1073/pnas.1200844109

Mayor, M., Bonfils, X., Forveille, T., et al. (2009) The HARPS search for southern extra-solar planets XVIII. An Earth-mass planet in the GJ 581 planetary system. *Astronomy & Astrophysics* 507:487-494. doi:10.1051/0004-6361/200912172

McKenna, D.D., Farrell, B.D. (2006) Tropical forests are both evolutionary cradles and museums of leaf beetle diversity. *Proceedings of the National Academy of Sciences* 103:10947-10951. doi:10.1073/pnas.0602712103

Miglio, A. and Montalbán, J. (2005) Constraining fundamental stellar parameters using seismology. Application to α Centauri AB. *Astronomy & Astrophysics* 441:615-629. doi:10.1051/0004-6361:20052988

Miguel, Y. and Brunini, A. (2010) Planet formation: statistics of spin rates and obliquities of extrasolar planets. *Monthly Notices of the Royal Astronomical Society* 406:1935-1943. doi:10.1111/j.1365-2966.2010.16804.x

Miller, S.R., Augustine, S., Olson, T.L., et al. (2004) Discovery of a free-living chlorophyll d-producing cyanobacterium with a hybrid proteobacterial/cyanobacterial small-subunit rRNA gene. *Proceedings of the National Academy of Sciences* 102:850-855. doi:10.1073/pnas.0405667102

Mojzsis, S.J., Arrhenius, G., McKeegan, K.D., Harrison, T.M., Nutman, A.P, Friend, C.R.L. (1996) Evidence for life on Earth before 3,800 million years ago. *Nature* 384:55-59. doi:10.1038/384055a0

Moreau, C.S., Bell, C.D. (2013) Testing the museum versus cradle tropical biological diversity hypothesis: Phylogeny, diversification, and ancestral biogeographic range evolution of the ants. *Evolution* 67:2240-2257. doi:10.1111/evo.12105

Muirhead, P.S., Johnson, J.A., Apps, K., et al. (2012) Characterizing the Cool KOIs. III. KOI 961: A Small Star with Large Proper Motion and Three Small Planets. *Astrophysical Journal* 747:144 (16 pp.). doi:10.1088/0004-637X/747/2/144

Noack, L. and Breuer, D. (2011) Plate Tectonics on Earth-like Planets. *EPSC Abstracts* 6:890-891. ADS:2011epsc.conf..890N

O'Brien, D.P., Geissler, P., and Greenberg, R. (2002) A Melt-Through Model for Chaos Formation on Europa. *Icarus* 156:152-161. doi:10.1006/icar.2001.6777

O'Malley-James, J.T., Greaves, J.S., Raven, J.A., Cockell, C.S. (2013) Swansong Biospheres: Refuges for life and novel microbial biospheres on terrestrial planets near the end of their habitable lifetimes. *International Journal of Astrobiology* 12:99-112. doi:10.1017/S147355041200047X

Ogihara, M. and Ida, S. (2012) N-body Simulations of Satellite Formation around Giant Planets: Origin of Orbital Configuration of the Galilean Moons. *Astrophysical Journal* 753:60 (17 pp.) doi:10.1088/0004-637X/753/1/60

Olson, P. and Christensen, U.R. (2006) Dipole moment scaling for convection-driven planetary dynamos. *Earth and Planetary Science Letters* 250:561-571. doi:10.1016/j.epsl.2006.08.008

Payne, J.L., McClain, C.R., Boyer, A.G., et al. (2011) The evolutionary consequences of oxygenic photosynthesis: a body size perspective. *Photosynthesis Research* 107:37-57. doi:10.1007/s11120-010-9593-1

Pérez, F. and Granger, B.E. (2007) IPython: a System for Interactive Scientific Computing. *Computing in Science and Engineering* 9:21-29. http://ipython.org, doi:10.1109/MCSE.2007.53

Pierrehumbert, R. (2010) Principles of Planetary Climate. Cambridge University Press. ISBN:9780521865562

Porter, S.B. and Grundy, W.M. (2011) Post-capture Evolution of Potentially Habitable Exomoons. *Astrophysical Journal Letters* 736:L14. doi:10.1088/2041-8205/736/1/L14

Rauer, H., Gebauer, S., von Paris, P., et al. (2011) Potential biosignatures in super-Earth atmospheres. I. Spectral appearance of super-Earths around M dwarfs. *Astronomy & Astrophysics* 529:A8. doi:10.1051/0004-6361/201014368

Raven, J. (2007) Photosynthesis in watercolors. *Nature* 448:418. doi:10.1038/448418a

Raymond, S.N., Quinn, T., Lunine, J.I. (2004) Making other earths: dynamical simulations of terrestrial planet formation and water delivery. *Icarus* 168:1-17. doi:10.1016/j.icarus.2003.11.019

Raymond, S.N., O'Brien, D.P., Morbidelli, A., and Kaib, N.A. (2009) Building the terrestrial planets: Constrained accretion in the inner Solar System. *Icarus* 203:644-662. doi:10.1016/j.icarus.2009.05.016

Remus, F., Mathis, S., Zahn, J.P., and Lainey, V. (2012) Anelastic tidal dissipation in multi-layer planets. *Astronomy & Astrophysics* 541:A165 (17 pp). doi:10.1051/0004-6361/201118595

Reynolds, R.T., McKay, C.P., Kasting, J.F. (1987) Europa, tidally heated oceans, and habitable zones around giant planets. *Advances in Space Research* 7:125-132. doi:10.1016/0273-1177(87)90364-4

Rushby, A.J., Claire, M.W., Osborn, H., and Watson, A.J. (2013) Habitable Zone Lifetimes of Exoplanets around Main Sequence Stars. *Astrobiology* 13:833-849. doi:10.1089/ast.2012.0938

Ryder, G. (2002) Mass flux in the ancient Earth-Moon system and benign implications for the origin of life on Earth. *Journal of Geophysical Research (Planets)* 107:6.1-6.13. doi:10.1029/2001JE001583

Sasaki, T., Stewart, G.R., Ida, S. (2010) Origin of the Different Architectures of the Jovian and Saturnian Satellite Systems. *Astrophysical Journal* 714:1052-1064. doi:10.1088/0004-637X/714/2/1052

Scalo, J., Kaltenegger, L., Segura, A.G., et al. (2007) M Stars as Targets for Terrestrial Exoplanet Searches And Biosignature Detection. *Astrobiology* 7:85-166. doi:10.1089/ast.2006.0125

Scharf, C.A. (2006) The Potential for Tidally Heated Icy and Temperate Moons around Exoplanets. *Astrophysical Journal* 648:1196-1205. doi:10.1086/505256

Schilling, G. (2007) Habitable, But Not Much Like Home. *Science* 316:528. doi:10.1126/science.316.5824.528b

Schopf, J.W. (1993) Microfossils of the Early Archean Apex Chert: New Evidence of the Antiquity of Life. *Science* 260:640-646. doi:10.1126/science.260.5108.640

Schopf, J.W. (2006) Fossils evidence of Archean life. *Science* 260:640-646. 10.1098/rstb.2006.1834

Selsis, F., Kasting, J.F., Levrard, B., et al. (2007) Habitable planets around the star Gliese 581? *Astronomy & Astrophysics* 476:1373-1387. doi:10.1051/0004-6361:20078091

Sohl, F., Sears, W.D., and Lorenz, R.D. (1995) Tidal dissipation on Titan. *Icarus* 115:278-294. doi:10.1006/icar.1995.1097

Spencer, J.R., Rathbun, J.A., Travis, L.D., et al. (2000) Io's Thermal Emission from the Galileo Photopolarimeter-Radiometer. *Science* 288:1198-1201. doi:10.1126/science.288.5469.1198

Spiegel, D.S., Menou, K., and Scharf, C.A. (2008) Habitable Climates. *Astrophysical Journal* 681:1609-1623. doi:10.1086/588089

Spiegel, D.S., Menou, K., Scharf, C.A. (2009) Habitable Climates: The Influence of Obliquity. *Astrophysical Journal* 691:596-610. doi:10.1088/0004-637X/691/1/596

Spiegel, D.S., Raymond, S.N., Dressing, C.D., et al. (2010) Generalized Milankovitch Cycles and Long-Term Climatic Habitability. *Astrophysical Journal* 721:1308-1318. doi:10.1088/0004-637X/721/2/1308

Stamenković, V., Breuer, D., and Spohn, T. (2011) Thermal and transport properties of mantle rock at high pressure: Applications to super-Earths. *Icarus* 216:572-596. doi:10.1016/j.icarus.2011.09.030

Stomp, M., Huisman, J., Stal, L.J., and Matthijs, H.C.P. (2007) Colorful niches of phototrophic microorganisms shaped by vibrations of the water molecule. *ISME Journal* 1:271-282. doi:10.1038/ismej.2007.59

Strughold, H. (1955) The medical problems of space flight. *International Record of Medicine* CLXVIII:570–575

Thévenin, F., Provost, J., Morel, P., et al. (2002) Asteroseismology and calibration of alpha Cen binary system. *Astronomy & Astrophysics* 392:L9-L12. doi:10.1051/0004-6361:20021074

Thoul, A., Scuflaire, R., Noels, A., et al. (2003) A new seismic analysis of Alpha Centauri. *Astronomy & Astrophysics* 402:293-297. doi:10.1051/0004-6361:20030244

von Paris, P., Selsis, F., Kitzmann, D., Rauer, H. (2013) The Dependence of the Ice-Albedo Feedback on Atmospheric Properties. *Astrobiology* 13:899-909. doi:10.1089/ast.2013.0993

Walker, J.C.G., Hays, P.B., and Kasting, J.F. (1981) A negative feedback mechanism for the long-term stabilization of the earth's surface temperature. *Journal of Geophysical Research* 86:9776-9782. doi:10.1029/JC086iC10p09776

Ward, P.D. and Brownlee, D. (2000). Rare Earth: Why Complex Life is Uncommon in the Universe. *Copernicus Books (Springer Verlag)*: ISBN 0-387-98701-0. ADS:2000rewc.book.....W

Weber, P. and Greenberg, J.M. (1985) Can spores survive in interstellar space? *Nature* 316:403-407. doi:10.1038/316403a0

Whewell, W. (1853) On the Plurality of Worlds. J. W. Parker and Son, London. ISBN:9780226894355

Wiegert, P.A. and Holman, M.J. (1997) The Stability of Planets in the Alpha Centauri System. *Astronomical Journal* 113:1445-1450. doi:10.1086/118360

Wignall, P.B. and Twitchett, R.J. (1996) Oceanic Anoxia and the End Permian Mass Extinction. *Science* 272:1155-1158. doi:10.1126/science.272.5265.1155

Williams, D.M. and Kasting, J.F. (1997a) Habitable Planets with High Obliquities. *Icarus* 129:254-267. doi:10.1006/icar.1997.5759

Williams, D.M., Kasting, J.F., and Wade, R.A. (1997b) Habitable moons around extrasolar planets. *Nature* 385:234-236. doi:10.1038/385234a0

Winn, J.N., Matthews, J.M., Dawson, R.I., et al. (2011) A Super-Earth Transiting a Naked-eye Star. *Astrophysical Journal* 737:L18 (6 pp.). doi:10.1088/2041-8205/737/1/L18

Wittgenstein, L. (1953) Philosophische Untersuchungen. Suhrkamp (2003). ISBN:978-3-518-22372-7

Wittenmyer, R.A., Tinney, C.G., Butler, R.P., et al. (2011) The Frequency of Low-mass Exoplanets. III. Toward $\eta_\oplus$ at Short Periods. *Astrophysical Journal* 738:81. doi:10.1088/0004-637X/738/1/81

Wordsworth, R. and Pierrehumbert, R. (2013) Water loss from terrestrial planets with $CO_2$-rich atmospheres. *Astrophysical Journal* 778:154. doi:10.1088/0004-637X/778/2/154

Yang, J., Cowan, N.B., and Abbot, D.S. (2013) Stabilizing Cloud Feedback Dramatically Expands the Habitable Zone of Tidally Locked Planets. *Astrophysical Journal* 771:L45 (6 pp). doi:10.1088/2041-8205/771/2/L45

Yoder, C.F. (1979) How tidal heating in Io drives the Galilean orbital resonance locks. *Nature* 279:767-770. doi:10.1038/279767a0

Zahnle, K., Arndt, N., Cockell, C., et al. (2007) Emergence of a Habitable Planet. *Space Science Reviews* 129:35-78. doi:10.1007/s11214-007-9225-z

No competing financial interests exist. This is a copy of an article published in Astrobiology 2014. Astrobiology is available online at http://online.liebertpub.com.